\documentclass{aa}

\usepackage[T1]{fontenc}

\usepackage[colorlinks=true,linkcolor=blue,citecolor=blue]{hyperref}

\usepackage{graphicx}
\usepackage{amsmath}
\usepackage{multicol,multirow}
\usepackage{tabularx}
\usepackage{booktabs} 
\usepackage{color,soul}
\usepackage{cleveref}
\usepackage{xspace} 
\usepackage{xcolor}
\usepackage{subcaption}
\usepackage{pdflscape}
\usepackage{placeins}
\usepackage{float}
\usepackage[flushleft]{threeparttable}

\usepackage{txfonts}

\usepackage{natbib}
\bibpunct{(}{)}{;}{a}{}{,}
\defcitealias{Gnarini.etAl.2025}{G25} 

\newcommand{\ixpe}{{IXPE}\xspace}
\newcommand{\nicer}{{NICER}\xspace }
\newcommand{\nustar}{\textit{NuSTAR}\xspace}

\newcommand{\fluxcgs}{erg~s$^{-1}$~cm$^{-2}$\xspace}
\newcommand{\lumcgs}{erg~s$^{-1}$\xspace}

\def\diskbb{\texttt{diskbb}\xspace}

\def\bbodyrad{\texttt{bbodyrad}\xspace}

\def\relxill{\texttt{relxill}\xspace}

\xspaceaddexceptions{+ - *}

\begin{document} 

\title{X-ray polarization of Z-type neutron star low-mass X-ray binaries}
\subtitle{II. Spectropolarimetric analysis}
\titlerunning{Polarization of Z-Sources II.}
\authorrunning{A. Gnarini et al.}

\author{Andrea Gnarini \inst{\ref{in:MSFC},\ref{in:UniRoma3}}
\and Francesco Ursini \inst{\ref{in:UniRoma3}}
\and Giorgio Matt \inst{\ref{in:UniRoma3}}
\and Stefano Bianchi \inst{\ref{in:UniRoma3}}
\and Fiamma Capitanio \inst{\ref{in:INAF-IAPS}}
\and Massimo Cocchi \inst{\ref{in:INAF-OAC}}
\and Sergio Fabiani \inst{\ref{in:INAF-IAPS}}
\and Ruben Farinelli \inst{\ref{in:INAF-OAS}}
\and Philip Kaaret \inst{\ref{in:MSFC}}
\and Lorenzo Marra \inst{\ref{in:INAF-IAPS}}
\and Antonella Tarana \inst{\ref{in:INAF-IAPS}}
}

\institute{NASA Marshall Space Flight Center, Huntsville, AL 35812, USA \label{in:MSFC}
\and
Dipartimento di Matematica e Fisica, Università degli Studi Roma Tre, via della Vasca Navale 84, I-00146 Roma, Italy \label{in:UniRoma3}
\and
INAF -- Istituto di Astrofisica e Planetologia Spaziali, Via del Fosso del Cavaliere 100, 00133 Roma, Italy \label{in:INAF-IAPS}
\and
INAF Osservatorio Astronomico di Cagliari, via della Scienza 5, I-09047 Selargius (CA), Italy \label{in:INAF-OAC}
\and
INAF -- Osservatorio di Astrofisica e Scienza dello Spazio, Via P. Gobetti 101, 40129 Bologna, Italy \label{in:INAF-OAS}
}

\date{Received XXX; accepted YYY}

\abstract{
    IXPE has provided for the first time detailed energy- and time-resolved X-ray polarimetry of Z-type neutron star low-mass X-ray binaries (NS-LMXBs) as they move along their color-color diagrams (CCDs). These sources can reach the highest polarization observed for NS-LXMBs in the 2--8 keV range when they move along the horizontal branch. In a previous paper, we characterized the spectral state of a sample of Z-sources using the CCD and estimated the polarization with model-independent analysis. Here, we present detailed spectropolarimetric analysis for each source on each branch using data from IXPE, NICER, and \textit{NuSTAR}. The continuum X-ray emission of all the sources is well described with a combination of thermal accretion disk emission plus a harder Comptonized component. In addition, reflection features, in particular the relativistically broadened Fe K$\alpha$ line, are observed for our sources, except GX~5--1. For most of the sources and branches, the main contribution to the X-ray emission and polarization is due to Comptonization: moving from the horizontal branch (HB) to the normal branch (NB), the polarization degree (PD) in the 2--8 keV band varies from about 6\% to 3--4\%, while the PD is loosely constrained in the flaring branch (FB), due to the shorter exposures. These PD values are significantly higher than theoretical expectations for typical spreading or boundary layer configurations. The polarization of the disk is generally lower (below 3\%) but still higher than predictions for an electron scattering-dominated, plane-parallel atmosphere above the disk observed at the corresponding inclination. Moreover, the polarization angle (PA) of the disk seems to be significantly misaligned and not perpendicular to that of Comptonization. We find no correlation between the polarization signal and the inclination, nor with the contribution of reflected photons throughout the Z-track.
}


\maketitle

\nolinenumbers

\section{Introduction}

Weakly magnetized neutron stars in low-mass X-ray binaries (NS-LMXBs) are systems characterized by a neutron star accreting matter through Roche-lobe overflow, typically from a late main-sequence or evolved degenerate companion star. The accretion flow is stopped by the surface of the NS, creating a boundary (BL) or spreading layer (SL) between the accretion disk and the NS surface. Z-sources are a class of NS-LMXBs with typical X-ray luminosity $L_{\rm X} > 10^{38}$ \lumcgs and exhibiting a three-branch track in their hardness-intensity (HID) or hard color-soft color diagrams (CCDs; \citealt{Hasinger.VanDerKlis.1989,VanDerKlis.1989}), characterized by the ``horizontal branch'' (HB), the ``normal branch'' (NB), and the ``flaring branch'' (FB). 

Typical X-ray spectra of NS-LMXBs are generally well described by soft thermal emission produced by the accretion disk \citep{Mitsuda.etAl.1984,Mitsuda.etAl.1989} or the NS surface \citep{White.etAl.1988} plus a harder component related to the inverse Compton scattering of soft photons in a hot electron plasma. However, the exact nature and geometry of the Comptonizing region are still a matter of debate since the different models are spectroscopically degenerate. In contrast, X-ray polarization strongly depends on the geometry of the system (e.g., \citealt{Gnarini.etAl.2022,Gnarini.etAl.2024,Farinelli.etAl.2025}) and is crucial to determining its physical characteristics. The reflection of X-ray photons above the surface of the accretion disk is another important component (see also the reviews by \citealt{DiSalvo.etAl.2024} and \citealt{Ludlam.2024}): reflection features have been detected in the spectra of most Z-sources \citep{Smale.1998,DAi.etAl.2009,Iaria.etAl.2009,Cackett.etAl.2010,Coughenour.etAl.2018,Mazzola.etAl.2021,Ludlam.etAl.2022,Thomas.etAl.2024}, although not in all (e.g., GX~5--1; \citealt{Homan.etAl.2018,Fabiani.etAl.2024}). These reflected photons are expected to be highly polarized \citep{Lapidus.Sunyaev.1985,Matt.1993}, significantly contributing to the polarization despite the fact that their relative contribution to the total X-ray emission may be quite low \citep{Ursini.etAl.2023,Gnarini.etAl.2024,Anitra.etAl.2025}.

X-ray polarimetry performed with the Imaging X-ray Polarimetry Explorer \citep[IXPE;][]{Weisskopf.2022} now represents a very powerful tool to study NS-LMXBs. In particular, the characteristics of the polarization signal are highly sensitive to the specific geometrical configuration and morphology of the Comptonizing region \citep{Gnarini.etAl.2022,Gnarini.etAl.2024,Farinelli.etAl.2024,Bobrikova.etAl.2025}. For example, a polarization lower than 2\% for a source observed at low inclination would rule out slab-like configurations, favoring spreading layer-like geometries; on the other hand, a high PD of 5--6\% would suggest a slab-shaped geometry at a high inclination angle. Being very bright sources, NS-LMXBs are perfect targets for X-ray polarimetry, and IXPE has detected significant polarization in most of them \citep[see][for a detailed review]{Ursini.2024}. 

In the first part of this work (\citealt{Gnarini.etAl.2025}, \citetalias{Gnarini.etAl.2025} hereafter), we performed model-independent analysis for each Z-source using \textsc{ixpeobssim} \citep{Baldini.etAl.2022} to measure polarization along the Z-track. We decided to restrict our sample to Z-sources for which IXPE data were publicly available before December 2024, namely Cyg~X-2, XTE~J1701--462, GX~5--1, Sco~X-1, GX~340+0, and GX~349+2. In this second part, we report the results of the first branch-resolved spectropolarimetric analysis of the sample of Z-sources observed with IXPE as they move along their CCDs. These sources exhibit the highest polarization in the full 2--8 keV \ixpe band when observed along the HB, up to about 4\% \citepalias{Gnarini.etAl.2025}; however, when they move along the NB, the polarization decreases ($\lesssim 2\%$) without any significant rotation of the polarization angle (PA). Although the FB was identified during the observations of all Z-sources, significant polarization was detected only for some of them \citepalias{Gnarini.etAl.2025}; in particular, the polarization seems to increase moving from the NB to the FB for Cyg~X-2 and Sco~X-1, while it remains consistent within the two branches at the 90\% confidence level for GX~5--1 and GX~349+2. 

Similarly to the first part of the work \citepalias{Gnarini.etAl.2025}, we considered the same procedure for data reduction and analysis for all Z-sources, to perform the spectropolarimetric analysis with \textsc{xspec} \citep{Arnaud.1996} using the same baseline model for a uniform spectral fitting. Using the same model for all sources allows us to make direct comparisons between the various Z-sources observed and to study the evolution of their physical parameters, for example, the temperature and optical depth of the Comptonizing region or the location of the inner edge of the accretion disk, as the sources move along their CCDs. The paper is structured as follows: in Sect. \ref{sec:Observations}, we report the data reduction to obtain the spectra for each observatory. In Sect. \ref{sec:Spec.Analysis}, we present the new X-ray spectropolarimetric analysis for each source; finally, in Sect. \ref{sec:Pol.Analysis} and \ref{sec:Conclusions}, we discuss the results obtained.

\section{Observations and data reduction}\label{sec:Observations}

All observations of the Z-sources with \ixpe, \nicer, and \nustar are introduced and listed in \citetalias{Gnarini.etAl.2025}. In addition to the data reduction procedure for each instrument already described in \citetalias{Gnarini.etAl.2025}, we extracted the spectra to analyze with \textsc{xspec} as follows.

\subsection{IXPE}

We extracted spectropolarimetric data for each detector unit (DU) using the standard dedicated \textsc{ftools} procedure with the latest available calibration files (CALDB v.20250225). The spectra for each Z-source have been extracted using the same source circular regions found in \citetalias{Gnarini.etAl.2025}, without applying any background rejection or subtraction \citep{DiMarco.etAl.2023}. The adopted radius of the extraction regions for each source and DU is derived using an iterative process to maximize the S/N in the 2--8 keV range. The resulting radii $R$ are reported in Table \ref{table:Obs}. Only for GX~5--1, we used a fixed 60$''$ radius region to avoid the X-ray halo observed with \textit{Chandra} \citep{Smith.etAl.2006,Clark.2018}. We applied the weighted analysis method described in \cite{DiMarco.etAl.2022} with the parameter \texttt{stokes=Neff} in \textsc{xselect}. For each DU, we produced the ancillary response file (ARF) and modulation response file (MRF) using the \texttt{ixpecalcarf} task, with the same source extraction radii. Then we rebinned all the \ixpe $I$ spectra using the standard \texttt{ftgrouppha} task and requiring a minimum S/N of 3 per bin. The Stokes $Q$ and $U$ spectra were then rebinned using the same energy binning as the $I$ spectra.

\renewcommand{\arraystretch}{1.1}
\begin{table}
\caption{Log of source extraction regions.}             
\label{table:Obs}      
\centering                                     
\begin{tabular}{l l c c c}         
\hline\hline       
\noalign{\smallskip}
  Satellite & Obs. ID & R.A. [h] & Decl. [deg] & $R$ [$''$] \\  
\hline     
\noalign{\smallskip}
\multicolumn{5}{c}{Cyg~X-2} \\
  \ixpe & 01001601 & 21:44:41.1 & +38:19:17.8 & 120 \\
  \ixpe & 01006601 & 21:44:41.7 & +38:19:11.7 & 120 \\
  \nustar & 30801012002 & 21:44:40.2 & +38:19:24.7 & 100 \\
  \hline
\multicolumn{5}{c}{XTE~J1071--461} \\
  \ixpe & 01250601 & 17:00:57.5 & --46:11:04.7 & 120 \\
  \ixpe & 01250701 & 17:00:57.3 & --46:11:01.1 & 120 \\
  \nustar & 90801325002 & 17:00:58.34 & --46:11:11.6 & 140 \\
  \hline
\multicolumn{5}{c}{GX~5--1$^{\star}$} \\
  \ixpe & 02002701 & 18:01:09.0 & --25:04:30.8 & 60 \\
  \ixpe & 02002702 & 18:01:09.2 & --25:04:31.0 & 60 \\
  \nustar & 90902310002 & 18:01:07.8 & --25:04:41.2 & 60 \\
  \nustar & 90902310004 & 18:01:07.9 & --25:04:42.4 & 60 \\
  \nustar & 90902310006 & 18:01:07.9 & --25:04:41.3 & 60 \\
  \hline
\multicolumn{5}{c}{Sco~X-1} \\
  \ixpe & 02002401 & 16:19:54.4 & --15:38:13.6 & 155 \\
  \nustar & 30902036002 & 16:19:55.3 & --15:38:25.5 & 180 \\
  \hline
\multicolumn{5}{c}{GX~340+0} \\
  \ixpe & 03003301 & 16:45:50.2 & --45:36:29.4 & 130 \\
  \ixpe & 03009901 & 16:45:46.7 & --45:36:35.2 & 130 \\
  \nustar & 91002313002 & 16:45:47.0 & --45:36:37.0 & 160 \\
  \nustar & 91002313004 & 16:45:46.9 & --45:36:36.8 & 160 \\
  \nustar & 30901012002 & 16:45:48.1 & --45:36:41.2 & 160 \\
  \hline
\multicolumn{5}{c}{GX~349+2} \\
  \ixpe & 03003601 & 17:05:43.7 & --36:25:13.6 & 120 \\
  \nustar & 91002333002 & 17:05:44.5 & --36:25:22.8 & 160 \\
  \nustar & 91002333004 & 17:05:44.6 & --36:25:23.2 & 160 \\
\hline      
\end{tabular}
\tablefoot{
$^{(\star)}$We considered a fixed 60$''$ radius region to avoid the X-ray halo detected with \textit{Chandra} \citep{Smith.etAl.2006,Clark.2018}.}
\end{table}

\subsection{NICER}

We derived \nicer spectra with the \texttt{nicerl3-spect} task available in HEASARC, while the background was computed using the SCORPEON\footnote{\url{https://heasarc.gsfc.nasa.gov/docs/nicer/analysis_threads/scorpeon-overview/}} model. The spectra were then automatically rebinned with the \texttt{ftgrouppha} task, considering the optimal binning algorithm by \cite{Kaastra.Bleeker.2016} with the additional requirement of 10 counts per grouped bin. In most of the \nicer spectra, we found some significant residuals, typically below 2 keV; these features are likely related to instrumental issues not considered in the \nicer ancillary response file (ARF; see also \citealt{Miller.etAl.2018,Strohmayer.etAl.2018}). Therefore, we added a multiplicative absorption edge (\texttt{edge}) to remove these features from the residuals. The resulting threshold energy of this edge obtained from the best-fits is very close to the Al edge at 1.839 keV. If we fix the threshold energy to the same value of the Al edge, the fits improve. For the spectral analysis of \nicer spectra, we adopted \texttt{PGstat} statistics since the background estimation for \nicer observations is done with SCORPEON and results in a non-Poissonian background.

\subsection{NuSTAR}

We extracted the spectra for each focal plane module (FPM) using the \texttt{nuproducts} command, with the same source and background circular regions found in \citetalias{Gnarini.etAl.2025}. Following the same approach adopted for IXPE, we derived the size of the extraction regions for each source and FPM by maximizing the S/N in the 3--79 keV range, while a fixed 60$''$ region is considered for the background. The resulting source radii $R$ for each source and observation are reported in Table \ref{table:Obs}. Similarly to the \ixpe spectra, the \nustar spectra were also rebinned using \texttt{ftgrouppha}, with the optimal binning algorithm by \cite{Kaastra.Bleeker.2016} and a minimum S/N of 3 per grouped bin. 

\section{Spectral analysis}\label{sec:Spec.Analysis}

\begin{figure*}[ht]
    \centering
    \begin{subfigure}[b]{0.325\textwidth}
    \centering
    \caption{}\label{fig:Spec.CygX-2}
    \includegraphics[width=\textwidth]{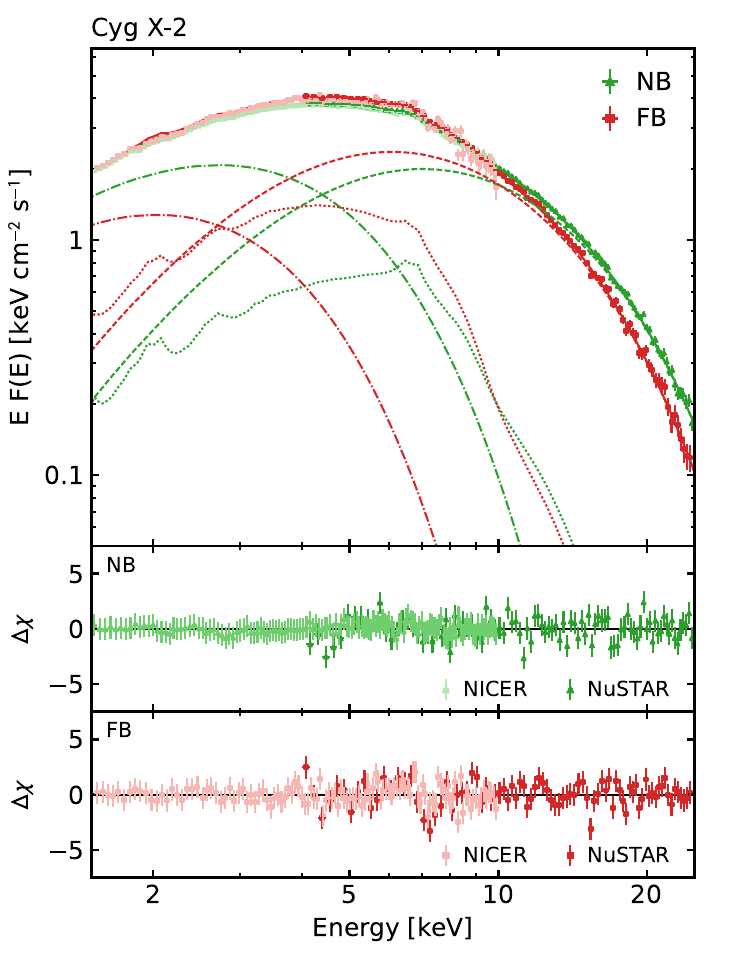}
    \end{subfigure}
    \begin{subfigure}[b]{0.325\textwidth}
    \centering
    \caption{}\label{fig:Spec.XTEJ1701}
    \includegraphics[width=\textwidth,height=8cm]{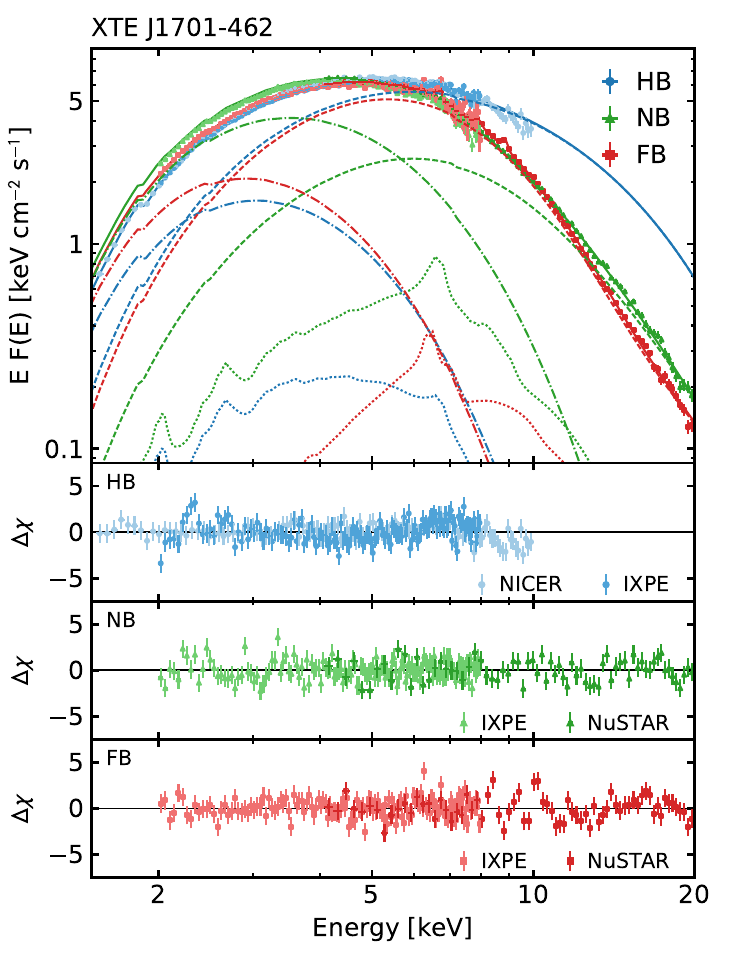}
    \end{subfigure}
    \begin{subfigure}[b]{0.325\textwidth}
    \centering
    \caption{}\label{fig:Spec.GX5-1}
    \includegraphics[width=\textwidth]{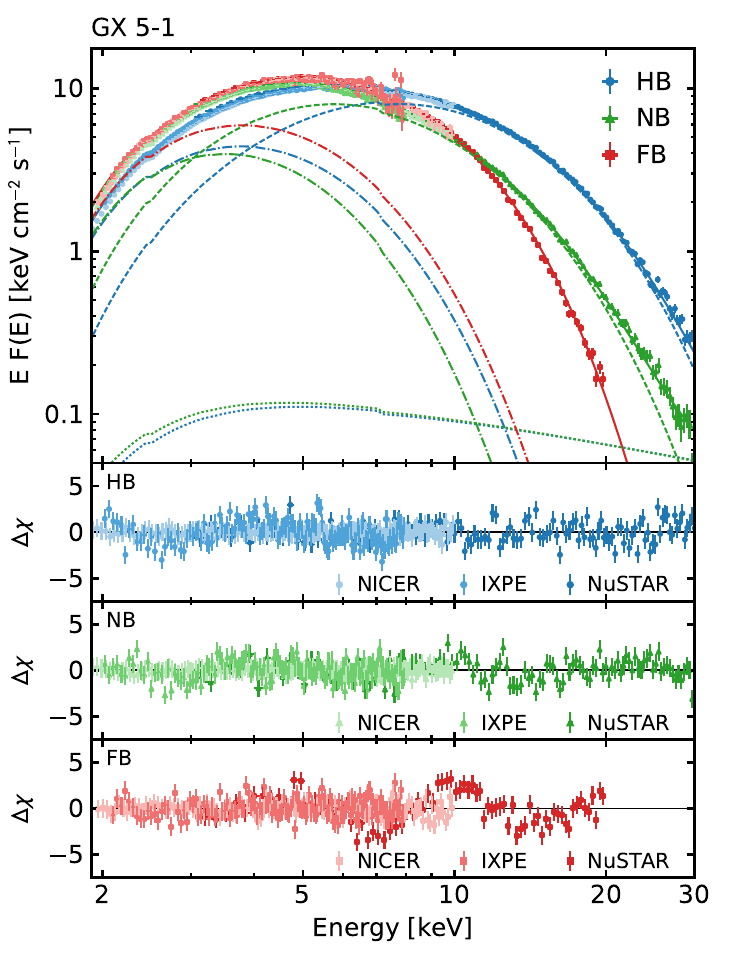}
    \end{subfigure}
    \\[1ex]
    \begin{subfigure}[b]{0.325\textwidth}
    \centering
    \caption{}\label{fig:Spec.ScoX-1}
    \includegraphics[width=\textwidth]{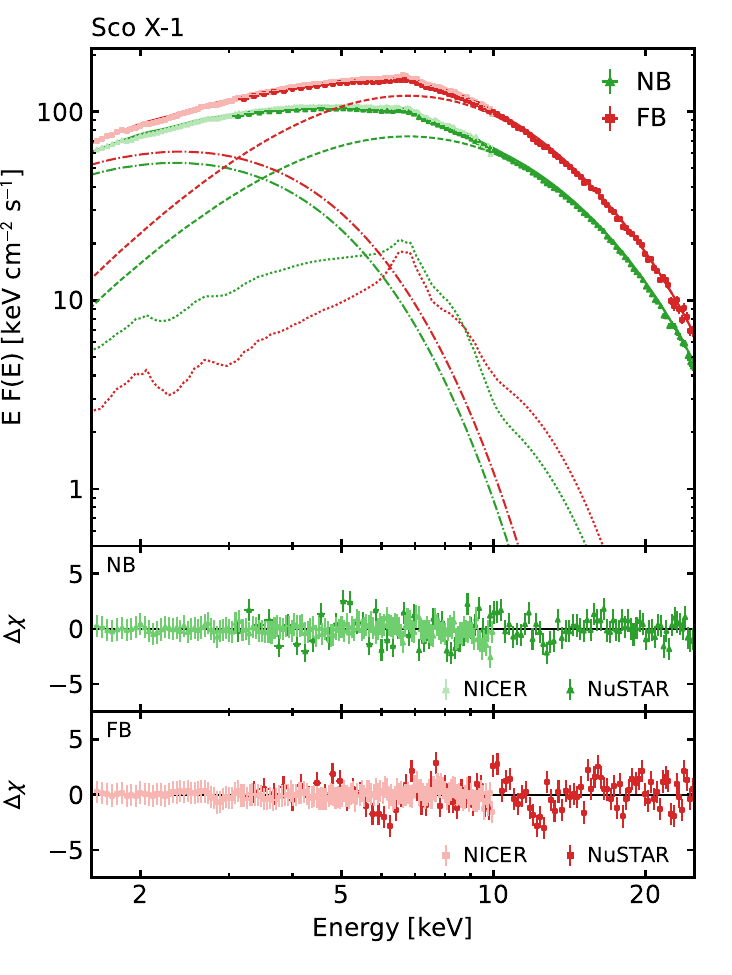}
    \end{subfigure}
    \begin{subfigure}[b]{0.325\textwidth}
    \centering
    \caption{}\label{fig:Spec.GX340+0}
    \includegraphics[width=\textwidth]{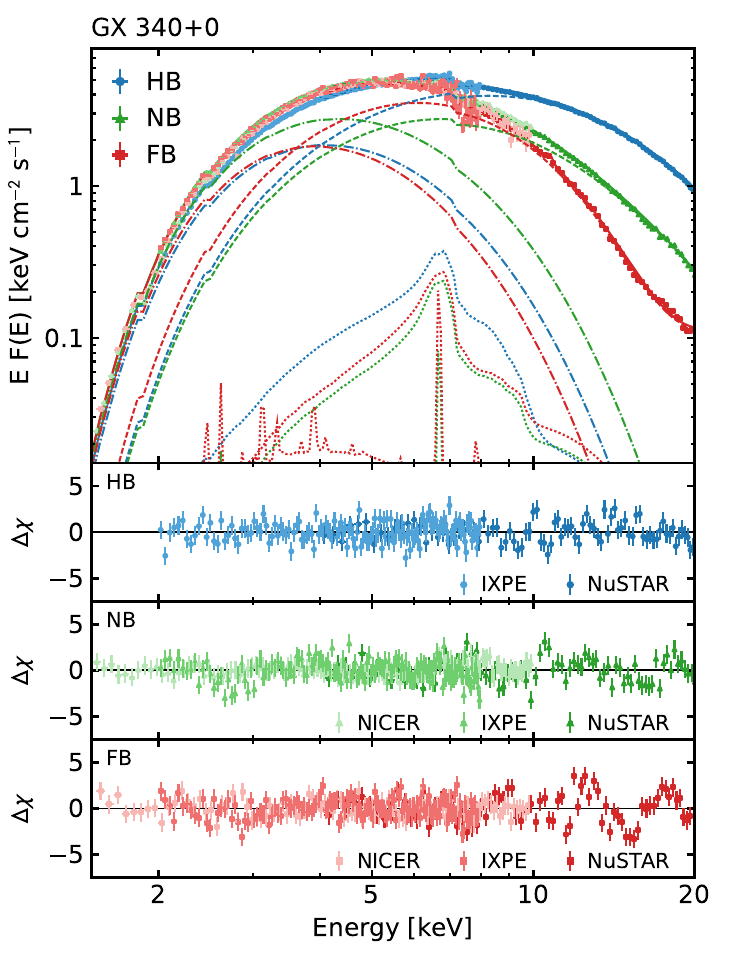}
    \end{subfigure}
    \begin{subfigure}[b]{0.325\textwidth}
    \centering
    \caption{}\label{fig:Spec.GX349+2}
    \includegraphics[width=\textwidth]{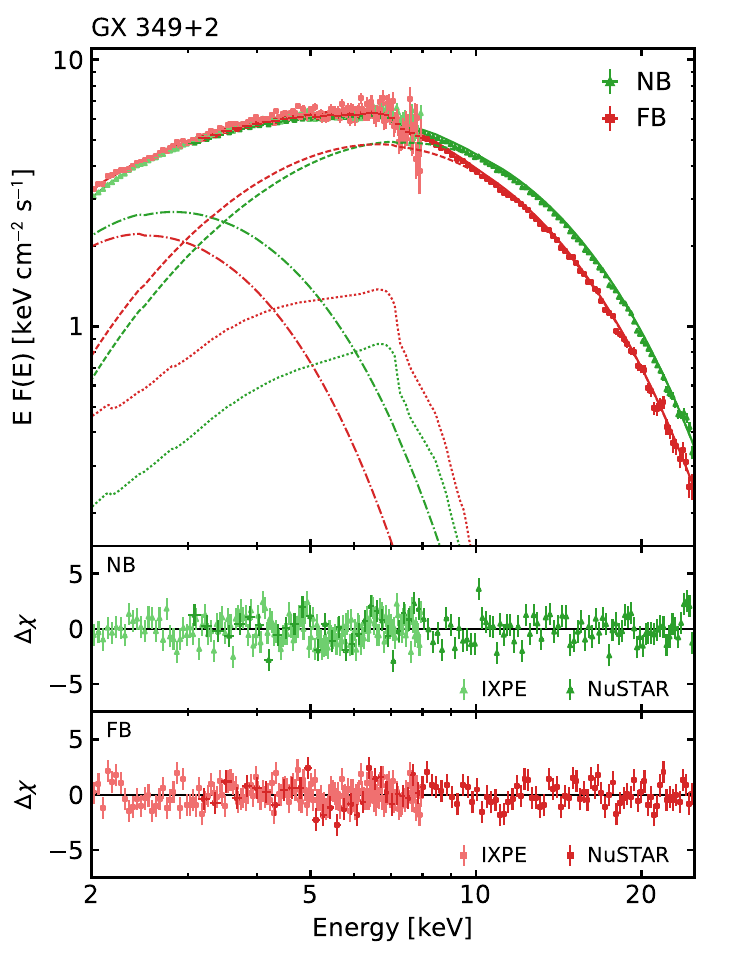}
    \end{subfigure}
    \caption{Deconvolved spectra for each Z-source with the resulting best-fit model and the corresponding residuals in units of $\sigma$. The three branches are highlighted with different colors and markers. Main spectral components are highlighted with different line styles, i.e., \texttt{diskbb} (dash-dotted lines), \texttt{thcomp*bbodyrad} (dashed lines), and \texttt{relxillNS} (dotted lines). (a) Cyg~X-2. (b) XTE~J1701–462. (c) GX~5–1. (d) Sco~X-1. (e) GX~340+0. (f) GX~349+2.}
    \label{fig:Spectra}
\end{figure*}

For each source, we performed a single joint fit of the \ixpe, \nustar, and \nicer spectra among the different branches of the Z-track with \textsc{xspec} \citep{Arnaud.1996}. The spectra are integrated along each branch and are derived using good time intervals (GTIs), as described in \citetalias{Gnarini.etAl.2025}. For all sources, we considered the \ixpe spectra in the 2--8 keV band, while for \nicer the 1.5--10 keV energy range is considered. For \nustar, the adopted energy ranges differ between sources and branches because the background starts to dominate at different energies, but typically in the 20--30 keV range. 

To have a direct comparison between the different Z-sources and a uniform spectral fitting, we decided to use the same two-component baseline model:
\begin{description}
\item {\tt TBabs*(diskbb+thcomp*bbodyrad)}.
\end{description}
An energy-independent cross-calibration multiplicative constant has been introduced for each IXPE DU, each \textit{NuSTAR} FPM and the NICER spectra. We modeled interstellar absorption using \texttt{tbabs}, considering \texttt{vern} cross section \citep{Verner.etAl.1996} and \texttt{wilm} abundances \citep{Wilms.etAl.2000}. The baseline model includes a multicolor disk blackbody (\texttt{diskbb}; \citealt{Mitsuda.etAl.1984}) and a harder Comptonized emission modeled with the convolution model \texttt{thcomp} \citep{Zdziarski.etAl.2020} applied to a \texttt{bbodyrad} component, associated with the NS emission. The covering fraction $f$ of \texttt{thcomp} represents the fraction of Comptonized photons with respect to the total seed photon emission.

All observed Z-sources, with the exception of GX~5--1 \citep[see also][]{Homan.etAl.2018,Fabiani.etAl.2024}, show evidence of an iron K$\alpha$ line between 6--7 keV and a reflection component. We employed the \texttt{relxillNS} model to account for the reflected photons. The \relxill models calculate the relativistic reflection from the innermost regions of the accretion disk \citep{Garcia.etAl.2014,Dauser.etAl.2014,Garcia.etAl.2022}; \texttt{relxillNS} is a particular flavor of \texttt{relxill} assuming a single-temperature blackbody spectrum as the primary continuum illuminating the accretion disk at 45\degr, physically related to the emission from the NS surface or from the boundary or spreading layer. In each branch, we decided to tie the temperature of the seed photons of \texttt{relxillNS} to the blackbody temperature of \texttt{bbodyrad}. The reflection fraction is set to $-1$ for all sources in order to obtain only the reflected emission without the direct continuum. The shape of the reflection spectrum is characterized by several physical parameters: the emissivity index $q_\text{em}$, the dimensionless spin $a$, the inner and outer radii of the accretion disk ($R_\text{in}$ and $R_\text{out}$ respectively), the inclination $i$ of the system, the number density $n_e$, the iron abundance $A_\text{Fe}$, and the ionization parameter $\xi$. For a standard NS, the dimensionless spin can be fixed to 0.1, consistent with the value that could be derived from the spin period obtained from quasi-periodic oscillations (\citealt{Wijnands.etAl.1998,Braje.etAl.2000}; see also \citealt{Patruno.2017} for a statistical analysis of the spin distributions of NS-LMXBs). Therefore, for all Z-sources, we fixed the value of $a$ to 0.1, while the outer disk radius is always fixed to 1000 gravitational radii. As the sources move along their Z-tracks, we expect that some parameters such as the inclination of the system, the iron abundance, and the emissivity index should remain constant differently, for example, from the inner disk edge.

\subsection{Cyg~X-2}

Cyg~X-2 was the first Z-source observed by \ixpe, simultaneously with \nicer and \nustar. The two observations were made early in the mission, before in-flight adjustments were made to establish a good alignment between the mirror module and overall telescope optical axes\footnote{\url{https://heasarc.gsfc.nasa.gov/FTP/ixpe/data/obs/01/01001601/README} and \url{https://heasarc.gsfc.nasa.gov/FTP/ixpe/data/obs/01/01006601/README}}. This is not modeled in the response matrices, thus for each DU the slope of the \ixpe spectra is slightly different and not compatible with those of \nicer and \nustar \cite[see also][]{Capitanio.etAl.2023,Farinelli.etAl.2023}. Therefore, spectral analysis was conducted using only these two observatories. It should be noted that since this issue impacts the Stokes parameters $I$, $Q$, and $U$ in the same way, the PD and PA remain unaffected. 

Cyg~X-2 was found mainly in the NB with a rapid excursion in the FB \citepalias{Gnarini.etAl.2025}. We left most of the physical parameters of the disk and the Comptonization component free to vary as the source moves along the different branches, along with the ionization parameter $\xi$ and the normalization of the reflection. The inclination $i$ and the emissivity index $q_\text{em}$ of \texttt{relxillNS} are tied between the branches but left free to vary, while the values of iron abundance $A_\text{Fe}$ and number density $n_{\rm e}$ were fixed to those obtained by \cite{Ludlam.etAl.2022}. In both branches, we found that the covering fraction $f$ is very close to 1, i.e., $>0.98$ and $>0.95$, for the NB and FB respectively; therefore, its value was fixed to 1 to better constrain the other physical parameters.

The deconvolved spectra with the resulting best-fit for each branch are reported in Fig. \ref{fig:Spec.CygX-2} and Table \ref{table:BestFit-CygX2}. The best-fit models obtained for the NB and FB provide statistically acceptable results, with $\chi^2/\text{d.o.f.}$ (degrees of freedom) of 350/309 and 318/275, respectively. The values of the disk and Comptonization parameters are consistent with the typical values for Cyg~X-2 \citep[][]{DiSalvo.etAl.2002,Farinelli.etAl.2009,Farinelli.etAl.2023}, while the results obtained for \texttt{relxillNS} are similar to those obtained by \cite{Ludlam.etAl.2022}. The radius of the blackbody-emitting region $R_{\rm bb}$ and the apparent inner radius of the disk $R_{\rm d}$ are derived from the normalizations of \texttt{bbodyrad} and \texttt{diskbb}, considering a distance to the source of 7.2 kpc \citep{Orosz.Kuulkers.1999}. As the source moves from the NB to the FB, both the \texttt{diskbb} and \texttt{bbodyrad} temperatures decrease. Additionally, the \texttt{thcomp} electron temperature decreases, whereas the optical depth remains almost unchanged. In the 2--8 keV range, accretion disk emission is the main contribution to the photon flux in the NB, while the Comptonization and the reflection are significantly more important in the FB. Recently, Cyg~X-2 was also observed in the HB by \cite{Gnarini.etAl.2026}. Because the spectropolarimetric analysis was performed with the same model and procedure, we can include those results in this analysis to obtain a complete coverage of the Z-track. In the HB, the total 2-8 keV flux is significantly dominated by Comptonized emission, characterized by an electron temperature slightly higher than the other branches but optical depth consistent within errors. On the other hand, the disk temperature is much lower than those of the NB and FB. 

\subsection{XTE~J1701--462}

XTE~J1701--462 was observed twice by \ixpe during its 2022 outburst \citep[see also][]{Cocchi.etAl.2023,Yu.etAl.2025}. In the first observation, simultaneously with \nicer, the source was identified in its HB, while in the second observation the source was tracking the NB and FB in its \nustar and \ixpe HIDs \citepalias{Gnarini.etAl.2025}. For all the branches, we left the physical parameters of \texttt{diskbb} and \texttt{thcomp*bbodyrad} free to vary. Reflection features were well identified in the \nustar spectra, while in the \nicer one there is a hint of the presence of the iron line between 6--7 keV. Therefore, we decided to characterize the properties of the reflection using the \nustar spectra of the second observation, leaving free to vary the inclination $i$, the inner disk radius $R_{\rm in}$, and the ionization parameter $\xi$. Then, we linked all the physical parameters of \texttt{relxillNS} for the first observation to that obtained for the second one but with only the normalization free to vary. 

The deconvolved spectra with the resulting best-fit for each branch are reported in Fig. \ref{fig:Spec.XTEJ1701} and Table \ref{table:BestFit-XTEJ1701}. The best-fit models obtained for the three branches yield statistically acceptable results, with $\chi^2/\text{d.o.f.}$ values of 565/552 for the HB, 615/575 for the NB, and 625/571 for the FB. Both the \texttt{diskbb} and the \texttt{bbodyrad} temperature varies moving from the HB to the FB, while the electron temperature of \texttt{thcomp} remains consistent within the errors. On the other hand, the covering fraction $f$ decreases significantly. In the HB, its value is always $>0.93$, therefore all the seed photons are Comptonized; in the NB only about 20\% of the blackbody photons result in Comptonization; in the FB, almost all the seed photons are observed since only about 2\% of them get scattered in the Comptonizing region. Moreover, the optical depth $\tau$ of \texttt{thcomp} increases as the source moves to the FB, up to about 30 keV, although its value in the FB is not well constrained because of the low fraction of Comptonized photons. From \texttt{relxillNS}, we found an inclination of about $30\degr$, consistent with the results obtained by \cite{Thomas.etAl.2024}, while only an upper limit for the inner disk radius is obtained in the HB and NB. In the FB, $R_{\rm in}$ results in being always unconstrained, and we decided to fix its value to the innermost stable circular orbit (ISCO)
radius. The ionization parameter significantly decreases moving to the FB. In the HB, the major contribution to the total 2--8 keV flux comes from Comptonized photons ($\approx 67\%$), while in the NB the disk contributes almost 70\% of the emission. The situation is a bit peculiar in the FB: although the main contribution comes from the \texttt{thcomp*bbodyrad} component, since the covering fraction is only 2\%, the 2--8 keV flux is dominated by the seed blackbody photons. For all the branches, the photon flux ratio of \texttt{thcomp*bbodyrad} significantly increases with energy, whereas the \texttt{diskbb} contribution decreases, as expected. 

\subsection{GX~5--1}

\ixpe observed GX~5--1 twice simultaneously with \nicer and \nustar in 2023. The source remained in the HB throughout the first observation, while it moved along the NB and FB during the second \citepalias{Gnarini.etAl.2025}. Unlike the other Z-sources, we do not detect any reflection features in the GX~5--1 spectra. Following \cite{Fabiani.etAl.2024}, we considered an additional hard tail component modeled with a powerlaw with a low-energy exponential rolloff (\texttt{expabs*powerlaw}; see also \citealt{Paizis.etAl.2006}). The energy of \texttt{expabs} is set equal to the temperature of \texttt{bbodyrad} since the powerlaw emission originates from the seed distribution of the Comptonized emission. The \texttt{powerlaw} photon index $\alpha$ is set to the best-fit value obtained by \cite{Fabiani.etAl.2024}, while normalization $N_{\rm pl}$ is considered a free parameter. In the FB, we do not observe any excess in the high-energy part of \nustar spectra: including the powerlaw, the fit statistic does not significantly improve and we obtain an upper limit of 0.02 for the normalization of \texttt{expabs*powerlaw}. Therefore, we removed this component in the FB. Along the different branches, we left the physical parameters of the disk and Comptonization free to vary during the fit, while we tied the value of the hydrogen column density $N_{\rm H}$. 

The deconvolved spectra with the resulting best-fit for each branch are reported in Fig. \ref{fig:Spec.GX5-1} and Table \ref{table:BestFit-GX5-1}. The best-fit models obtained for the three branches yield statistically acceptable results, with $\chi^2/\text{d.o.f.}$ values of 976/881 for the HB, 906/814 for the NB, and 871/708 for the FB. During the first observation with the source in the HB, the covering fraction $f$ result is $>0.97$, therefore we decided to fix its value to 1. For the second observation, differently from \cite{Fabiani.etAl.2024} in which the best-fit was performed using time-averaged spectra with a best-fit value for $f$ ranging between 2.7\% and 11.2\%, we obtained a covering fraction $>0.98$ and $<0.01$ for NB and FB, respectively. Similarly to the HB, we fixed the value of $f$ to 1 for the NB to better constrain the other physical parameters of \texttt{thcomp}. For the FB, we noticed that the fit improves by fixing the covering fraction to 0 or removing the \texttt{thcomp} component. We tried also to apply \texttt{thcomp} to the disk emission to verify if the fit would improve; in this case also the covering fraction result is $<0.02$ and the fit is statistically very similar to that with \texttt{thcomp*bbodyrad}. Therefore, in the FB, the spectra are well modeled by a combination of the disk emission and a simple blackbody. As the source moves from the HB to the NB, both the temperature of the disk and of the seed blackbody component decrease, while they increase again in the FB. In addition, the temperature and the optical depth of the Comptonizing region also decrease from the HB to the NB. Unlike Cyg~X-2 and XTE~J1701--462, for GX~5--1, the main contribution to the total 2--8 keV flux comes from the disk emission for both the HB and the FB (up to 60\%), while only in the NB does Comptonization represent the major component. On the other hand, the contribution of \texttt{thcomp*bbodyrad}, or only of \texttt{bbodyrad} in the FB, increases significantly with energy for all branches, while that of \texttt{diskbb} decreases.

\subsection{Sco~X-1}

Sco~X-1 was observed in 2023 by IXPE using the gray filter to reduce its X-ray flux, simultaneously with both \nicer and \nustar \citep{LaMonaca.etAl.2024}. In particular, \nustar identified the source moving between the NB and the FB during \ixpe exposure \citepalias{Gnarini.etAl.2025}. For the spectral analysis, we considered only the \nicer and \nustar spectra, since the \ixpe spectra exhibit significant residuals in the entire 2--8 keV energy band, likely related to the use of the gray filter. To correct for this issue, during the polarimetric analysis, in addition to the cross-calibration constant, we multiplied the model by a powerlaw difference ($E^{-\Delta \Gamma}$), following a similar procedure to \cite{Ludlam.etAl.2022}. 

The deconvolved spectra with the resulting best-fit for each branch are reported in Fig. \ref{fig:Spec.ScoX-1} and Table \ref{table:BestFit-ScoX1}. The best-fit models obtained for the NB and FB provide statistically acceptable results, with $\chi^2/\text{d.o.f.}$ of 440/367 and 474/367, respectively. The physical parameters of the disk and Comptonized emission are left free to vary during the fit, as long as the inclination $i$, the emissivity index $q_{\rm em}$, the iron abundance $A_{\rm Fe}$, and the ionization parameter $\xi$ of \texttt{relxillNS}. The values of the disk and electron temperature are consistent within errors between the NB and FB, while the optical depth and the temperature of \texttt{bbodyrad} significantly increases in the FB. Since in the NB the covering fraction is very close to 1, i.e., $>0.99$, its value is fixed to 1 to find better constraints for the other physical parameters. However, in the FB, the covering fraction $f$ ranges between 48\% and 54\%; therefore, not all the seed photons of \texttt{bbodyrad} are Comptonized. The parameters of the Comptonized component are also consistent with the results obtained by \cite{Mazzola.etAl.2021}, while the \texttt{diskbb} parameters are slightly different, probably due to the different value of $N_{\rm H}$. Unlike \cite{LaMonaca.etAl.2024}, we left the inclination free to vary and its measured value ($46\degr \pm 3\degr$) is consistent with the results obtained by \cite{Fomalont.etAl.2001,Fomalont.etAl.2001.b}, while we were able to constrain the value of the inner disk radius in the NB ($\approx 2.2$ $R_{\rm ISCO}$), but only an upper limit of 1.8 $R_{\rm ISCO}$ is found for the FB. These results for the inner disk radius represent an improvement with respect to the upper limit obtained by \cite{LaMonaca.etAl.2024}, while the best-fit values for the ionization parameters are slightly different, probably as a result of the different spectral model and the assumptions for the reflected component. It should also be noted that the spectral analysis in \cite{LaMonaca.etAl.2024} was performed using time-averaged spectra, rather than fitting the different branches separately. In the 2--8 keV energy band, the main contribution to the total flux comes from the Comptonization in both branches, about 45\% in the NB and 55\% in the FB, whereas the contribution of both the disk and reflection decreases from the NB to the FB. For both branches, both the photon flux ratios of \texttt{thcomp*bbodyrad} and \texttt{relxillNS} increase with energy, while the contribution of \texttt{diskbb} decreases from about 55\% to less than 10\%. 

\renewcommand{\arraystretch}{1.15}

\begin{table*}[t]
\caption{Polarization degree and angle of each spectral component for each source with the corresponding branch of the Z-track.} 
\label{table:Pol}      
\centering                                     
\begin{tabular}{lc cc cc cc}
\hline\hline       
\noalign{\smallskip}
\multirow{2}{*}{Source} & \multirow{2}{*}{Branch} & \multicolumn{2}{c}{\texttt{diskbb}} & \multicolumn{2}{c}{\texttt{thcomp*bbodyrad}}  & \multicolumn{2}{c}{\texttt{relxillNS}}  \\
 & & PD [\%] & PA [deg] & PD [\%] & PA [deg] & PD [\%] & PA [deg] \\
\hline     
\noalign{\smallskip}
Cyg~X-2 & HB$^{\ast}$ & $<2.6$ & = PA$_\text{Comp}+90$ & $4.2 \pm 0.9$ & $-52 \pm 3$ & [10] & = PA$_\text{Comp}$ \\
Cyg~X-2 & NB & $2.4 \pm 1.1$ & 54 $\pm$ 15 & 3.5 $\pm$ 1.7 & $-42 \pm 7$ & [10] & = PA$_\text{Comp}$ \\
Cyg~X-2 & FB & [2.3] & = PA$_\text{Comp}+90$ & $<5.2$ & $-44 \pm 12$ & [10] & = PA$_\text{Comp}$ \\\hline
XTE~J1701--462 & HB & $3.8 \pm 1.4$ & $13 \pm 11$ & $6.7 \pm 1.1$ & $-44 \pm 5$ & [10] & = PA$_\text{Comp}$ \\
XTE~J1701--462 & NB & $2.3 \pm 0.8$ & $32 \pm 13$ & $5.6 \pm 0.8$ & $-63 \pm 9$ & [10] & = PA$_\text{Comp}$ \\
XTE~J1701--462 & FB & $<4.1$ & = PA$_\text{Comp}+90$ & $1.6 \pm 1.1$ & $-37 \pm 10$ & [10] & = PA$_\text{Comp}$ \\\hline
GX~5--1 & HB & $2.3 \pm 1.3$ & $14 \pm 17$ & $6.4 \pm 1.4$ & $-18 \pm 6$ & - & - \\
GX~5--1 & NB & $2.8 \pm 1.9$ & $25 \pm 12$ & $2.9 \pm 1.4$ & $-30 \pm 8$ & - & - \\
GX~5--1 & FB & [2] & $29 \pm 19$ & $5.4 \pm 2.2$ & $-19 \pm 12$ & - & - \\\hline
Sco~X-1 & NB & $1.8 \pm 1.0$ & $-64 \pm 16$ & $<0.7$ & $9 \pm 10$ & [10] & = PA$_\text{Comp}$ \\
Sco~X-1 & FB & $<2.5$ & = PA$_\text{Comp}+90$ & $<1.9$ & $16 \pm 8$ & [10] & = PA$_\text{Comp}$ \\\hline
GX~340+0 & HB & $2.9 \pm 1.5$ & $7 \pm 15$ & $6.0 \pm 1.6$ & $48 \pm 7$ & [10] & = PA$_\text{Comp}$ \\
GX~340+0 & NB & $<1.6$ & = PA$_\text{Comp}+90$ & $3.4 \pm 1.5$ & $42 \pm 7$ & [10] & = PA$_\text{Comp}$ \\
GX~340+0 & FB & $<8.9$ & = PA$_\text{Comp}+90$ & $6.0 \pm 4.6$ & $45 \pm 19$ & [10] & = PA$_\text{Comp}$ \\\hline
GX~349+2 & NB & $2.3 \pm 0.8$ & $-8 \pm 15$ & $<1.3$ & $67 \pm 9$ & [10] & = PA$_\text{Comp}$ \\
GX~349+2 & FB & $<7.6$ & - & $<1.1$ & - & [10] & = PA$_\text{Comp}$ \\
\hline
\hline                                            
\end{tabular}
\tablefoot{Errors are quoted at the 90\% confidence level for a single parameter, while upper limits are reported at the 99\% confidence level. The parameters in square brackets were kept frozen during the fit. $^{(\ast)}$Results taken by \cite{Gnarini.etAl.2026}.}
\end{table*}

\subsection{GX~340+0}

IXPE observed GX~340+0 twice during the first cycle of the Guest Observing (GO) program in 2024 \citep[see also][]{Bhargava.etAl.2024.O1,Bhargava.etAl.2024.O2,LaMonaca.etAl.2025.GX340+0.O1,LaMonaca.etAl.2025.GX340+0.O2}. During the second IXPE observation,  \nicer also simultaneously observed the the source; while no \nustar observation was strictly simultaneous to the IXPE ones, they were performed a few days after the IXPE ones. For each branch, we verified that the \nustar spectra were consistent with those of \ixpe and, therefore, can be used for spectral modeling. To account for some residuals in the \nustar spectra, in addition to the baseline model and the \texttt{relxillNS} component, we applied the same ionized emission plasma model \texttt{apec} used by \cite{Ludlam.etAl.2025}, which analyzed the same \nustar spectra along with XRISM data. Since the parameters of \texttt{apec} are not constrained using only \nustar, \nicer, and \ixpe, we decided to fix the plasma temperature and the metal abundances to the same value reported by \cite{Ludlam.etAl.2025}. Similarly, we also fixed the value of the emissivity index $q_{\rm em}$ and the iron abundance $A_{\rm Fe}$ to the results obtained with XRISM.

The deconvolved spectra with the resulting best-fit for each branch are reported in Fig. \ref{fig:Spec.GX340+0} and Table \ref{table:BestFit-GX340+0}. The best-fit models obtained for the three branches yield statistically acceptable results, with $\chi^2/\text{d.o.f.}$ values of 763/675 for the HB, 808/750 for the NB, and 785/682 for the FB. Although the temperatures of \texttt{bbodyrad} and \texttt{diskbb} remain consistent within errors as the source moves along the Z-track, there is a clear evolution of the physical parameters of \texttt{thcomp}. In the HB, the covering fraction $f$ is $>0.99$ and we fixed its value to 1 to better constrain the other parameters. In the NB, the covering fraction decreases, ranging between 30\% and 40\%, while it reduces again in the FB, with only about 1\% Comptonized photons with respect to total blackbody emission. The optical depth $\tau$ increases significantly from HB to FB, while the electron temperature $kT_{\rm e}$ is consistent within the errors between the HB and the NB; in the FB, the electron temperature reaches about 15 keV, which is much higher than the results obtained for other Z-sources. However, the results for the FB of GX~340+0 are poorly constrained with respect to the other branches or other Z-sources, likely related to the very small fraction of Comptonized photons. The inclination $i$ can be well constrained and results are consistent with the results obtained by \cite{Ludlam.etAl.2025}, while only upper limits are obtained for the inner disk radius. For each branch, the main contribution to the 2--8 keV flux is the direct accretion disk emission, about 50\% in the HB and FB, and up to 67\% in the NB. Comptonization contributes roughly to about 45\% of the flux in the HB and FB, but only about 30\% in the NB. As for the other Z-sources, for all branches, the contribution of \texttt{diskbb} significantly decreases with energy, whereas that of \texttt{thcomp*bbodyrad} and \texttt{relxillNS} increases. 

\subsection{GX~349+2}

GX~349+2 was observed for the first time by \ixpe in September 2024 during the first GO cycle \citep[see also][]{Kashyap.etAl.2025.GX349+2,LaMonaca.etAl.2025.GX349+2}. \nustar also observed the source twice simultaneously with \ixpe, indicating that it was moving between the NB and the FB \citep{Gnarini.etAl.2025}. All the physical parameters of disk and Comptonization are left free to vary between the branches, while we tied the hydrogen column density of \texttt{TBabs}. Similarly, the inclination $i$ and the iron abundance $A_{\rm Fe}$ are also tied between the NB and FB, and we fixed the emissivity index $q_{\rm em}$ and the number density to a reasonable value to better constrain the other parameters. The inclination of the source can be well constrained with the \nustar data, while only an upper limit can be obtained for the inner disk radius in the NB. Due to the short exposure of the FB, the inner disk radius is unconstrained and we fixed its value to 1 $R_{\rm ISCO}$.

The deconvolved spectra with the resulting best-fit for each branch are reported in Fig. \ref{fig:Spec.GX349+2} and Table \ref{table:BestFit-GX349+2}. The best-fit models obtained for the NB and FB provide statistically acceptable results, with $\chi^2/\text{d.o.f.}$ of 683/647 and 657/632, respectively. Moving from the NB to the FB, the temperatures of the \texttt{diskbb} and \texttt{bbodyrad} decrease, while the temperature and the optical depth of the Comptonizing region results are consistent within errors between the branches. In both branches, the covering fraction $f$ is very close to 1, i.e., $>0.99$ and $>0.97$, for the NB and FB respectively; therefore, we decided to fix its value to 1 to better constrain the other parameters. The best-fit values for the physical parameters of \texttt{diskbb} and \texttt{thcomp*bbodyrad} differ slightly from those reported by \cite{Kashyap.etAl.2025.GX349+2} and \cite{LaMonaca.etAl.2025.GX349+2}, likely as a consequence of the different model adopted, but they are in line with the typical values of Z-type NS-LMXBs. From the reflection component, we found an inclination of $37\degr \pm 5\degr$, which is consistent with the results obtained by \cite{Coughenour.etAl.2018} and \cite{LaMonaca.etAl.2025.GX349+2}. However, we obtained only an upper limit in the NB for the inner disk radius of 1.6 $R_{\rm ISCO}$, while it is not constrained in the FB and its value was fixed at 1 $R_{\rm ISCO}$. The ionization parameter is consistent within the errors between the two branches. In both NB and FB, the main contribution to the total 2--8 keV flux is due to Comptonized photons (about 50\%). The contribution of the disk decreases from about 44\% to 33\% moving from the NB to the FB, whereas the photon flux ratio of reflection increases from 9\% to 17\%. Similarly to the other sources, we found an increasing contribution of \texttt{thcomp*bbodyrad} and \texttt{relxillNS} with energy, while a decreasing behavior with energy is observed for the contribution of \texttt{diskbb}. 

\section{Spectropolarimetric analysis}\label{sec:Pol.Analysis}

Once we obtained the best-fit models for each source and each branch of the Z-track, we performed the spectropolarimetric analysis fixing all spectral parameters to their best-fit values and using only the IXPE Stokes spectra. We first applied the multiplicative \texttt{polconst} component, describing a constant polarization with energy, to the entire best-fit models. The results for the polarization are consistent with those obtained using \texttt{PCUBE} \citepalias{Gnarini.etAl.2025}. We then switched \texttt{polconst} with the multiplicative \texttt{pollin} model in order to test the linear dependence of the polarization with energy. In this case, we also recovered the same results obtained with the linear fit applied to the \texttt{PCUBE} results \citepalias{Gnarini.etAl.2025}. Then, we applied the \texttt{polconst} component separately to each spectral component. Unlike previous spectropolarimetric studies of Z-sources \citep{Farinelli.etAl.2023,Cocchi.etAl.2023,LaMonaca.etAl.2024,LaMonaca.etAl.2025.GX340+0.O2,Kashyap.etAl.2025.GX349+2}, in which either a full reflection model was not used or reflection was not appropriately included in the spectropolarimetric analysis, we included the contribution of reflection to the polarization, as reflected photons can be highly polarized, up to about 10\% or 20\%, depending on the geometry and the ionization state of the disk \citep{Matt.1993,Podgorny.etAl.2025}. However, it is difficult to estimate the PD and PA for each component due to the limited bandpass of IXPE and the degeneracy of some components; for example, the contribution of the Comptonized and reflected radiation is not well constrained without some theoretical and observational expectations, since they peak at similar energies. 

For all Z-sources and branches, we decided to fix the PD of \texttt{relxillNS} at 10\% \citep{Matt.1993,Poutanen.Svensson.1996}, while the PA is tied to that of \texttt{thcomp*bbodyrad}. This assumption is reasonable for a typical configuration with an optically thick boundary or spreading layer characterized by $H \gg \Delta R$, where $H$ is its vertical height and $\Delta R$ its radial extension \citep{Poutanen.Svensson.1996,Schnittman.Krolik.2009}. For most sources, the reflection fraction is relatively small; therefore, fixing the PD does not significantly influence the results. In some cases, following the same assumption, we also decided to fix the value of the PA of \texttt{diskbb} to be perpendicular to that of \texttt{thcomp*bbodyrad} and \texttt{relxillNS} to obtain a constraint for the PD of the disk component. For GX~5--1, since the powerlaw component is related to the same seed distribution as the Comptonized component, we also linked the polarimetric parameters of \texttt{expabs*powerlaw} to those of \texttt{thcomp*bbodyrad}. For sources in which one or more Gaussian components were included to model an emission line, the polarization of those components was fixed to zero.

\begin{figure}
    \centering
    \includegraphics[width=0.975\linewidth]{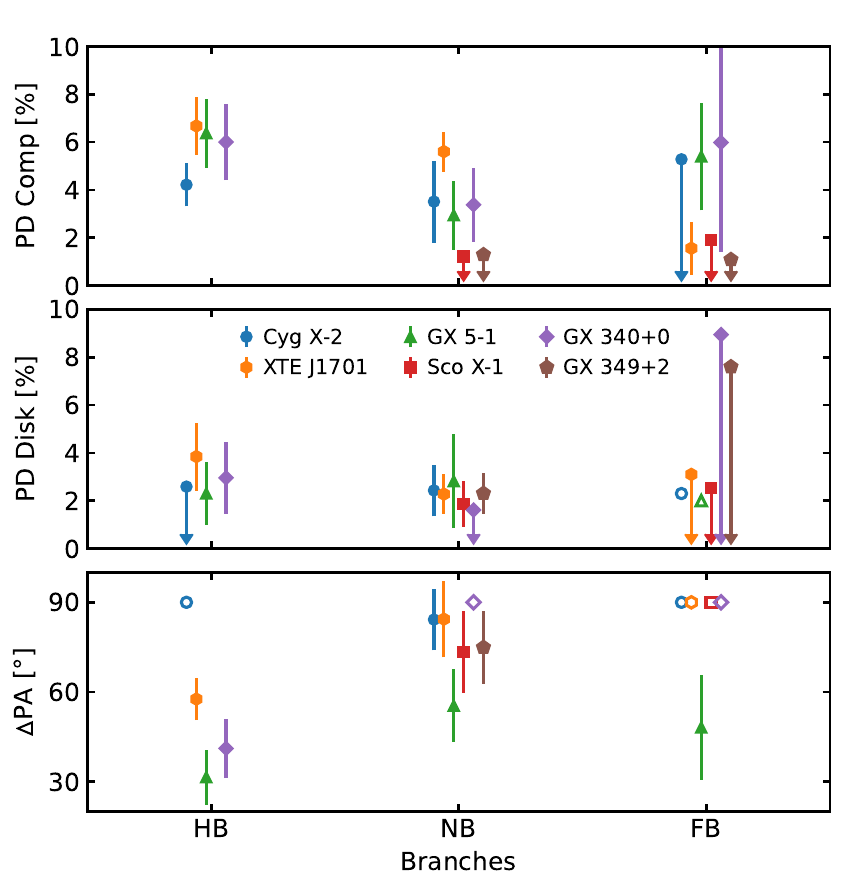}
    \caption{Polarization degree of Comptonized (top) and disk (middle) emission, and the difference of the polarization angle of the two components ($\Delta {\rm PA}$; bottom) for each Z-source as a function of the branch (see Table \ref{table:Pol}). Empty markers correspond to values frozen during the fits. Errors are at the 90\% confidence level; upper limits are reported at the 99\% confidence level for one interesting parameter.}
    \label{fig:Pol.Comp}
\end{figure}

The results of the spectropolarimetric analysis applying \texttt{polconst} to each spectral component are reported in Table \ref{table:Pol} and Fig. \ref{fig:Pol.Comp}. Compared with previous studies reporting spectropolarimetric results for Z-sources \citep{Farinelli.etAl.2023,Cocchi.etAl.2023,Fabiani.etAl.2024,LaMonaca.etAl.2024,LaMonaca.etAl.2025.GX340+0.O1,LaMonaca.etAl.2025.GX340+0.O2,LaMonaca.etAl.2025.GX349+2}, we performed the analysis by fitting the three individual branches separately. A similar approach was adopted only by \cite{Kashyap.etAl.2025.GX349+2} for GX~349+2, although using a three-component model that includes direct disk and NS emission, plus Comptonized disk photons, and without adopting a full reflection model. The resulting PD of the disk is generally lower with respect to the results obtained for Comptonization and reflection, with typical values ranging between 2--3\%. Only for Cyg~X-2 are the measured values consistent with the ``classical'' results for an electron scattering-dominated semi-infinite plane-parallel atmosphere \citep{Chandrasekhar.1960,Sobolev.1963} observed at the specific inclination derived from \texttt{relxillNS}. However, for all other sources, the disk polarization obtained from spectropolarimetric analysis is higher compared to the PD calculated by \cite{Chandrasekhar.1960} considering the measured inclination. A PD exceeding that of a pure-scattering atmosphere can be achieved when accounting for absorption and considering an atmospheric ionization far from complete \citep{Taverna.etAl.2021,Marra.etAl.2026}. When measurable, in most cases, the PA of the disk emission is significantly misaligned and not perpendicular to that of Comptonization, as expected for a typical geometrically thin but optically thick accretion disk. 

\begin{figure}
    \centering
    \includegraphics[width=0.975\linewidth]{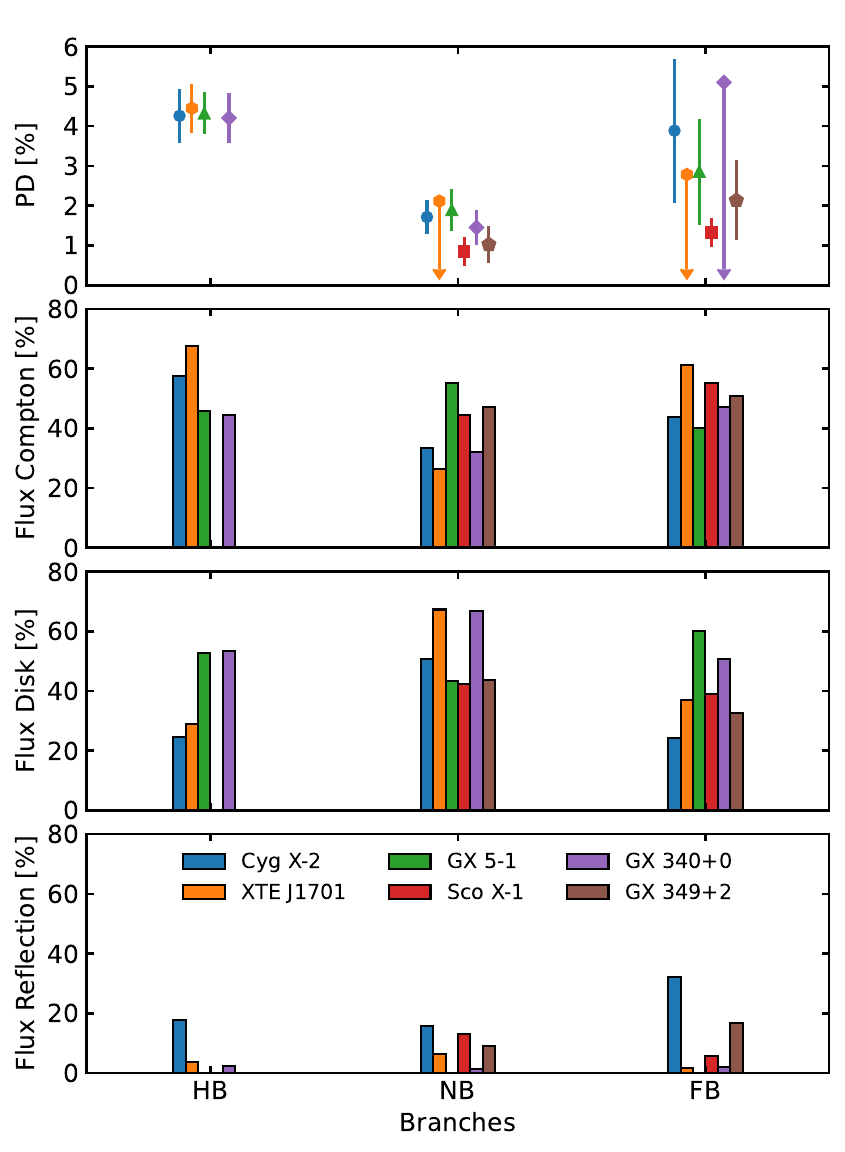}
    \caption{Average polarization in the 2--8 keV range (top) and photon flux ratios of Comptonized, disk, and reflected emission for each Z-source as a function of the branch. Errors are at the 90\% confidence level; upper limits are reported at the 99\% confidence level for one interesting parameter.}
    \label{fig:Flux.Comp}
\end{figure}

\begin{figure*}
    \centering
    \includegraphics[width=0.475\linewidth]{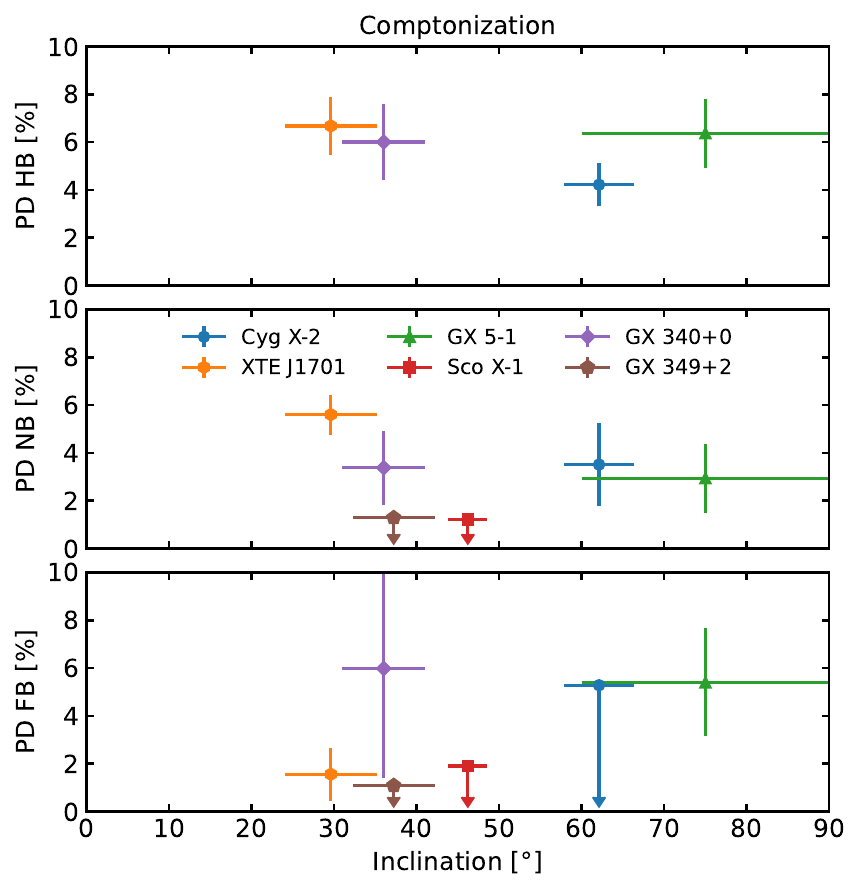}
    \includegraphics[width=0.475\linewidth]{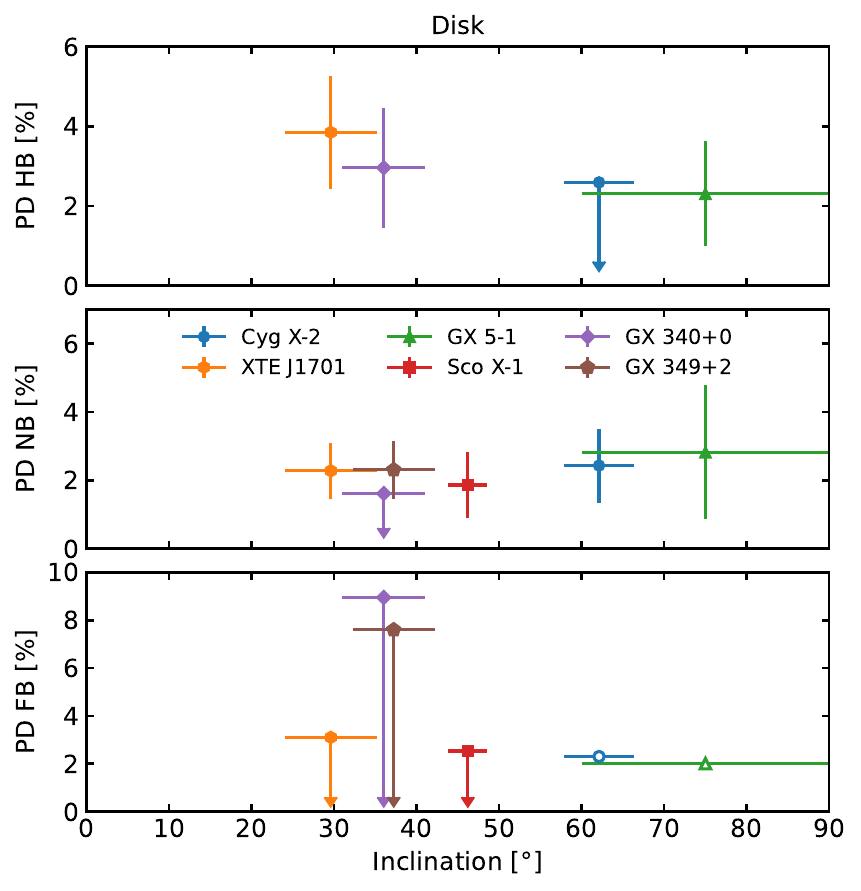}
    \caption{Polarization degree of Comptonized (left) and disk (right) emission for each Z-source as a function of the inclination. Empty markers correspond to values frozen during the fits. Errors are at the 90\% confidence level; upper limits are reported at the 99\% confidence level for one interesting parameter.}
    \label{fig:Pol.Incl}
\end{figure*}

In contrast, the Comptonization component exhibits a systematically higher PD, with values also reaching about 6\% in the HB, up to $6.7\% \pm 1.4\%$ for XTE~J1701--462. For GX~5--1 and GX~340+0, the PD of \texttt{thcomp*bbodyrad} in the NB is lower with respect to the HB values, while the PD is consistent within the errors between the HB and the NB for Cyg~X-2 and XTE~J1701--462. Despite the results in the FB exhibiting larger errors or upper limits due to the shorter exposures of the FB intervals, there is a hint of a scenario with variations in the geometry of the Comptonizing region throughout the Z-track providing a significant difference in the average 2--8 keV PD but without strong rotation of the PA. Typical spreading layer configurations are characterized by a polarization lower than about 2\% \citep{Gnarini.etAl.2022,Gnarini.etAl.2024,Farinelli.etAl.2024,Bobrikova.etAl.2025}, while the results obtained for \texttt{thcomp*bbodyrad} from spectropolarimetric analysis are generally significantly higher. The only exceptions are XTE~J1701--462 (FB), Sco~X-1 (NB and FB), and GX~349+2 (NB and FB), for which the PD of \texttt{thcomp*bbodyrad} is less than 2\% and consistent with the results for an optically thick spreading layer. Therefore, for other sources and branches, the Comptonizing region is expected to differ from typical spreading layer geometries to reproduce the high PD measured. For example, if we consider a boundary layer configuration characterized by a small opening angle, covering only the equatorial part of the NS surface, but which extends more radially from the NS to the inner edge of the accretion disk, the geometry of the Comptonizing region is less spherically symmetric, and the resulting PD could be higher than typical spreading layers. Nevertheless, it must be noted that there is a clear variation in the covering fraction parameter $f$ along the Z-track for XTE~J1701--462, GX~5--1, Sco~X-1, and GX~340+0 (see Tables \ref{table:BestFit-XTEJ1701}, \ref{table:BestFit-GX5-1}, \ref{table:BestFit-ScoX1}, and \ref{table:BestFit-GX340+0}); therefore, the measured PD for \texttt{thcomp*bbodyrad} represents the total contribution to the polarization of Comptonized photons plus the fraction of seed blackbody photons, which are expected to be unpolarized. For all Z-sources, the PA of \texttt{thcomp*bbodyrad} is consistent with the value derived with the model-independent analysis in the 2--8 keV band \citepalias{Gnarini.etAl.2025}, as expected, being the main contribution to the polarized signal. Therefore, considering Cyg~X-2, the PA of Comptonization is still aligned with the direction of the radio jet \citep{Spencer.etAl.2013}, while for Sco~X-1 results rotated by approximately $40\degr$ \citep{Fomalont.etAl.2001,Fomalont.etAl.2001.b,Long.etAl.2022}.

In Fig. \ref{fig:Flux.Comp}, we report the average PD derived with the model-independent analysis described in \citetalias{Gnarini.etAl.2025} and the photon flux ratios for the Comptonized, disk, and reflected emission for each Z-source as a function of the branch (see also Tables \ref{table:BestFit-CygX2}, \ref{table:BestFit-XTEJ1701}, \ref{table:BestFit-GX5-1}, \ref{table:BestFit-ScoX1}, \ref{table:BestFit-GX340+0}, and \ref{table:BestFit-GX349+2}). The contribution of the disk seems to be more important in the NB with respect to the other two branches, whereas the photon flux ratio of reflection is generally lower than 10\% for most of the sources and branches. Although no evident dependence between the polarization and the contribution of reflected photons among different branches can be claimed, there is a hint of correlation with the other two components: while the photon flux ratios of Comptonization seem to decrease from the HB to the NB, the contribution of the disk generally increases, and the polarization decreases. Therefore, the different contribution of the spectral components seems to be one of the mechanisms responsible for the variation of the polarization throughout the CCD. Completing the sample of Z-sources observed in each branch would be crucial to confirm this behavior.

In Fig. \ref{fig:Pol.Incl}, we also reported the PD of the Comptonized and disk emission as a function of the inclination derived from the reflection component during spectral analysis for each source and each branch. For GX~5--1, no reflection features are observed \citep[see also][]{Homan.etAl.2018,Fabiani.etAl.2024}; however, the absence of these features may be due to a highly ionized accretion disk with low iron abundance observed at high inclination ($\gtrsim 60\degr$; \citealt{Kuulkers.etAl.1994,Homan.etAl.2018}). For the Comptonization, the results do not show any significant correlation between the polarization and the inclination. As for the disk emission, the PD is relatively stable with the inclination for the HB and the NB with values consistent within the errors, whereas in the FB only upper limits are found. Differing from the expectation of an increasing polarization with inclination, some sources at intermediate inclinations ($\approx 30-40\degr$, i.e., XTE~J1701--462 and GX~340+0) are characterized by a PD comparable to or even higher than other sources at higher inclination. These results may suggest a possible break in the axial symmetry of these systems: if the configuration is not axially symmetric, the dependence of the PD with inclination may differ from the increasing trend. In addition, the disk and Comptonized components would not be orthogonally polarized, as observed for some Z-sources (e.g., XTE~J1701--462, GX~5--1, and GX~340+0). We note that it may be possible that the value of the inclination derived during the best-fits with \texttt{relxillNS} could be underestimated: Cyg~X-2 is the only source exhibiting an inclination higher than 50$\degr$ and the result is consistent with previous measurements obtained with the same methodology \citep{Ludlam.etAl.2022} and using optical observations \citep{Orosz.Kuulkers.1999}; for other Z-sources, the inclination is lower, but typically consistent with the results obtained from the spectral fitting of the reflection component. Sco~X-1 is the only Z-source considered for which the inclination obtained with \texttt{relxillNS} is in perfect agreement with the results obtained from radio observation with very long baseline interferometry (VLBI) by \cite{Fomalont.etAl.2001,Fomalont.etAl.2001.b}. However, no other estimates of the inclination using different methodologies are available for the other sources. Therefore, we cannot exclude the possibility that some of these Z-sources could be observed at higher inclinations than the measured ones, although the absence of eclipses suggests that their inclinations are not extremely high. 

In addition, we did not find any significant differences between Cyg~X-2-like and Sco~X-1-like sources; considering the average PD in the 2--8 keV energy band (Fig. \ref{fig:Flux.Comp}; see also \citetalias{Gnarini.etAl.2025}), the values for Cyg~X-2-like and Sco~X-1-like sources are very similar, but we cannot exclude any differences in the HB since no Sco~X-1-like sources were observed in this branch. From the spectropolarimetric analysis of the different spectral components (Fig. \ref{fig:Pol.Comp}), we notice that in the NB and FB, only upper limits are obtained for the PD of Comptonization of Sco~X-1-like sources, while for the disk the results are always consistent for every source and branch. Although every source is observed in the FB, the polarization is typically less constrained compared to the other branches, and we cannot draw robust conclusions about the two subclasses; since the FB exhibits markedly different properties in Cyg~X-2-like and Sco~X-1-like sources \citep[see][]{Church.etAl.2012}, improved constraints on the polarization in this branch could provide valuable insight into the physical processes underlying this difference between the two types of Z-sources. Recently, \cite{Kashyap.etAl.2025.GX17+2} reported the first polarimetric results for the Sco~X-1-like Z-source GX~17+2: the X-ray emission of GX 17+2 is characterized by a PD = 1.9\% $\pm$ 0.3\% and PA = 11$\degr$ $\pm$ 4$\degr$ as the source moves along the NB. The measured PD is in line with other Z-sources in the NB, and both the PD and PA are consistent with the radio polarization measured with the Very Large Array at 10 GHz \citep{Kashyap.etAl.2025.GX17+2}. Similarly to other Z-sources discussed in this work, the spectra of GX 17+2 are well described by a combination of soft accretion disk emission, a harder Comptonized component, and reflection off the accretion disk surface. Further observations of the source in the HB and FB are crucial in determining whether its behavior is consistent with that of other Z-sources.

\section{Conclusions}\label{sec:Conclusions}

In this work, we performed a new branch-resolved spectropolarimetric analysis of the Z-sources observed by IXPE, after the energy- and branch-resolved model-independent polarimetric analysis with \textsc{ixpeobssim} reported by \citetalias{Gnarini.etAl.2025}. In particular, differently from previous spectral analyzes, we employed the same model for all sources to directly compare the spectropolarimetric results. Comparing the best-fit obtained for different Z-sources along each branch, we do not observe any significant trends for the physical parameters of \texttt{diskbb} and \texttt{thcomp*bbodyrad}; see also Tables \ref{table:BestFit-CygX2}, \ref{table:BestFit-XTEJ1701}, \ref{table:BestFit-GX5-1}, \ref{table:BestFit-ScoX1}, \ref{table:BestFit-GX340+0}, and \ref{table:BestFit-GX349+2}. The best-fit values for the election temperature and optical depth of the Comptonizing region are very similar for all sources and branches, except for XTE~J1701--462 and GX~340+0 in the FB, which are characterized by a covering fraction $f$ of \texttt{thcomp*bbodyrad} very close to zero, which also leads to higher uncertainties for $kT_{\rm e}$ and $\tau$. In contrast, the physical parameters of the disk and the blackbody seed photons vary moderately across different branches, although not in the same way for all sources. Consequently, the relative contributions of the two spectral components to the total photon flux differ significantly (see Fig. \ref{fig:Flux.Comp}). As in the case of the inclination, no clear or significant correlation is found between the observed polarization and any physical parameter of \texttt{diskbb} or \texttt{thcomp*bbodyrad}. In most of the cases, Comptonization represents the main contribution to the total 2--8 keV flux, whereas disk emission typically dominates at lower energies ($\lesssim 4$ keV). Although the contribution of reflection generally ranges between 5\% and 20\%, its contribution to polarization is not negligible, since reflected photons are expected to be highly polarized \citep{Matt.1993,Poutanen.Svensson.1996}. The only exception is GX~5--1 for which we have not observed any reflection features \citep[see also][]{Homan.etAl.2018,Fabiani.etAl.2024}. However, we note that the polarimetric properties of this source do not differ significantly from the other Z-sources, suggesting that reflection does not have a dominant role in any case. A detailed study on the contribution of reflection was performed by \cite{Liu.etAl.2026} for Cyg~X-2, also adopting different reflection models during  spectral analysis but without resolving the different branches. Although there is strong degeneracy between the Comptonized and reflected components, as expected, the upper limits derived for the reflection are consistent with the assumption that we adopted for each Z-source.

The disk emission results are generally lower polarized compared to the harder components but only for Cyg~X-2 is the resulting PD consistent with the prediction for a plane-parallel atmosphere above the accretion disk dominated by electron scatterings \citep{Chandrasekhar.1960,Sobolev.1963} for the corresponding inclination obtained from the best-fit models, while for other sources the PD is generally higher than expected. The PD of Comptonized photons varies between branches: for most sources, its values are higher than expected for typical spreading or boundary layer configurations \citep{Farinelli.etAl.2024,Bobrikova.etAl.2025}. Such high polarization levels would necessitate low optical depths, which is inconsistent with the properties of Z-sources, or with different geometries of the Comptonizing region. This leaves open the possibility of an additional component contributing to the polarized signal, despite not influencing the continuum spectrum significantly enough to be appreciably detectable through current spectroscopy, such as a sub-relativistic wind intercepting a significant fraction of the Comptonized radiation \citep{Rogantini.etAl.2025} or a completely ionized wind. We want to remark that we cannot exclude the possibility that disk photons are also Comptonized, as in the case of extended accretion disk corona systems \citep[ADC; e.g.,][]{White.Holt.1982,Parmar.White.1988,Miller.etAl.2016,Ludlam.etAl.2025,Mizumoto.etAl.2026}. However, this configuration would be characterized by a PD that is typically higher than the measured values for Atolls and Z-sources \citep[e.g.,][]{Gnarini.etAl.2024}.

The average polarization, as well as the polarization of the different components, seems to be correlated neither with the inclination, as could be expected for axially symmetric systems, nor with the variations of the reflected contribution throughout the CCD. On the other hand, the contribution of the disk is generally higher moving from the HB to the NB \citep[see also][]{Psaltis.etAl.1995}; since the PA of the disk emission is misaligned with that of the Comptonization component, an increase in the flux of the disk inevitably leads to a decrease in the net polarization. As reported in Fig. \ref{fig:Pol.Comp}, the PAs of the two components are not necessarily orthogonal: for some Z-sources, they differ by as little as 30$\degr$. Furthermore, for statistical reasons, the PA of reflection was tied to that of the Comptonized emission. Although this assumption holds for a strictly axis-symmetric geometry, a warped disk would require the reflection component to rotate, following the local surface normal to the disk. We also note the difficulty in drawing definitive conclusions about FB, mostly due to short exposure. Indeed, only for XTE~J1701--462 is a clear trend observed, where the PD associated with the Comptonized component progressively decreases along the Z-track, while for the remaining sources, we obtain either upper limits only or associated uncertainties that are too large to be conclusive. Consequently, a complete mapping of the polarimetric behavior of Z-sources in the 2--8 keV band along the entire track would be achievable with the upcoming eXTP mission \citep{Zhou.etAl.2025}.

Although the properties of Z-sources are not yet fully understood, IXPE is offering unprecedented insight into this class of NS-LMXBs. Compared to Atolls, Z-sources are significantly more polarized in the HB, while in the NB their PD is similar to that measured for Atoll sources ($\lesssim 2\%$; \citealt{Capitanio.etAl.2023,Ursini.etAl.2023,DiMarco.etAl.2023.4U,Ursini.etAl.2024,Gnarini.etAl.2024.GX3+1,Tarana.etAl.2025}). Moreover, for both classes of NS-LMXBs, the polarization increases with energy, while the PA remains typically constant. On the other hand, recently IXPE observed two highly inclined NS-LXMBs exhibiting the highest PD measured so far. The transient source AX~J1745.6--2901 \citep{Mikusincova.etAl.2025} was serendipitously detected during the observation of MAXI~J1744--294 \citep{Marra.etAl.2025} with a PD of $14.7\% \pm 4.0\%$, obtained after a detailed analysis of  contamination due to the Galactic center and the MAXI~J1744--294 emission; 2S~0921--630 (=V395 Car; \citealt{Tomaru.etAl.2025}) is characterized by a polarization of $8.8\% \pm 1.4\%$ and PD increasing with energy. These sources are typical ADC systems, in which the outer vertically extended region of the accretion disk obscures the central region of the system due to the high inclination ($\gtrsim 80\degr$; \citealt{Ashcraft.etAl.2012,Ponti.etAl.2015,Ponti.etAl.2018}). For these sources, the extremely high PD seems to be related to the presence of an accretion disk wind. Therefore, a similar effect applied to Z-sources observed at lower inclinations may reproduce the high PD measured in the HB but having a lower polarization for the Comptonized emission. These results on Z-sources emphasize the need for improved spectropolarimetric models, in order to move beyond the purely phenomenological approach, which is crucial also for future polarimetric missions with higher sensitivity and broader energy coverage than IXPE. An initial step in this direction was provided by \cite{Farinelli.etAl.2025} with a spectropolarimetric tabular model based on Monte Carlo simulations for high-soft state NS-LMXBs, including Comptonization within the SL and direct and disk-reflected emission, and by \cite{Podgorny.etAl.2025} with a detailed spectropolarimetric model for the reflected thermal emission, taking into account scattering, absorption, and spectral lines for a slab in photoionization equilibrium above the accretion disk. These improved theoretical models, supported by new X-ray spectropolarimetric observations, will provide an answer to understanding the geometry of these accreting sources. 


\begin{acknowledgements}
AG was supported by an appointment to the NASA Postdoctoral Program at the Marshall Space Flight Center (MSFC), administered by Oak Ridge Associated Universities under contract with NASA. This research was supported by the Italian Space Agency (Agenzia Spaziale Italiana, ASI) through the contract ASI-INAF-2022-19-HH.0. AT, FC, and SF acknowledge financial support by the Istituto Nazionale di Astrofisica (INAF) grant 1.05.24.02.04: ``A multi frequency spectro-polarimetric campaign to explore spin and geometry in Low Mass X-ray Binaries''. SF and LM have been supported by the project PRIN 2022 - 2022LWPEXW - ``An X-ray view of compact objects in polarized light'', CUP C53D23001180006. This work reports observations obtained with the Imaging X-ray Polarimetry Explorer (IXPE), a joint US (NASA) and Italian (ASI) mission, led by MSFC. The research uses data products provided by the IXPE Science Operations Center (MSFC), using algorithms developed by the IXPE Collaboration (MSFC, Istituto Nazionale di Astrofisica - INAF, Istituto Nazionale di Fisica Nucleare - INFN, ASI Space Science Data Center - SSDC), and distributed by the High-Energy Astrophysics Science Archive Research Center (HEASARC). This research has made use of data from the \textit{NuSTAR} mission, a project led by the California Institute of Technology, managed by the Jet Propulsion Laboratory, and funded by NASA. Data analysis was performed using the \textit{NuSTAR} Data Analysis Software (NuSTARDAS), jointly developed by the ASI Science Data Center (SSDC, Italy) and the California Institute of Technology (USA).
\end{acknowledgements}

\bibliographystyle{aa} 
\bibliography{References.bib} 

\begin{appendix}
\section{Best-fit tables}\label{sec:Appendix}

\renewcommand{\arraystretch}{1.25}

\begin{table}[H]
\caption{Best-fitting model parameters for Cyg~X-2.} 
\label{table:BestFit-CygX2}      
\centering                         
\small
\begin{tabular}{@{}llccc@{}}
\noalign{\smallskip}
\hline\hline         
& Parameter & HB$^\star$ & NB & FB \\   
\hline     
\noalign{\smallskip}
\texttt{TBabs} & $N_{\rm H}$ ($10^{22}$\,cm$^{-2}$) & [0.12] & \multicolumn{2}{c}{0.16$^{+0.05}_{-0.05}$} \\
\texttt{edge} & $E_{\rm c}$ (keV) & / & [1.839] & [1.839] \\
 & $D$ ($10^{-2}$) & / & 3.9$^{+0.2}_{-0.2}$ & 5.4$^{+0.6}_{-0.5}$ \\
\texttt{diskbb} & $kT_\text{in}$ (keV) & 0.65$^{+0.02}_{-0.02}$ & 1.12$^{+0.03}_{-0.03}$ & 0.81$^{+0.04}_{-0.04}$ \\
 & $R_{\rm d} \sqrt{\cos i}$ (km) & 23.3$^{+0.8}_{-0.8}$ & 10.2$^{+0.5}_{-0.5}$ & 15.7$^{+1.2}_{-1.2}$ \\
 \texttt{thcomp} & $kT_{\rm e}$ (keV) & 3.4$^{+0.1}_{-0.1}$ & 3.1$^{+0.1}_{-0.1}$ & 2.9$^{+0.1}_{-0.1}$ \\
 & $\tau$ & 8.5$^{+0.5}_{-0.4}$ & 9.0$^{+0.6}_{-0.7}$ & 8.8$^{+0.8}_{-0.8}$ \\
 & $f$ & [1] & [1] & [1] \\
\texttt{bbodyrad} & $kT$ (keV) & 1.10$^{+0.01}_{-0.01}$ & 1.28$^{+0.02}_{-0.02}$ & 1.13$^{+0.02}_{-0.02}$ \\
& $R_\text{bb}$ (km) & 11.6$^{+0.3}_{-0.3}$ & 8.2$^{+0.5}_{-0.4}$ & 11.5$^{+0.6}_{-0.6}$ \\
\texttt{relxillNS} & $q_\text{em}$ & 1.6$^{+0.2}_{-0.2}$ & \multicolumn{2}{c}{1.7$^{+0.2}_{-0.2}$} \\
& $a$ & & [0.1] & \\
& $i$ (deg) & [62] & \multicolumn{2}{c}{62$^{+4}_{-4}$} \\
& $R_\text{in}$ ($R_{\rm ISCO}$) & $< 5.3$ & $< 1.6$& [1] \\
& $kT_\text{bb}$ (keV) & = $kT$ & = $kT$ & = $kT$ \\
& $\log \xi$ & 3.0$^{+0.1}_{-0.1}$ & 2.6$^{+0.1}_{-0.1}$ & 2.8$^{+0.1}_{-0.1}$ \\
& $A_\text{Fe}$ & [1.4] & \multicolumn{2}{c}{[1.4]} \\
& $\log n_{\rm e}$ (cm$^{-3}$) & [18] & \multicolumn{2}{c}{[18]} \\
& $N_{\rm r}$ ($10^{-3}$) & 2.8$^{+0.8}_{-0.9}$ & 5.5$^{+0.2}_{-0.2}$ & 8.5$^{+0.2}_{-0.1}$ \\
\hline
\multicolumn{5}{c}{Cross-calibration} \\
  \multicolumn{2}{l}{$\mathcal{C}_\text{DU2/DU1}$} & / & 1.001$^{+0.001}_{-0.001}$ & 1.006$^{+0.005}_{-0.004}$ \\
  \multicolumn{2}{l}{$\mathcal{C}_\text{DU3/DU1}$} & 0.970$^{+0.003}_{-0.002}$ & 0.976$^{+0.002}_{-0.001}$ & 1.020$^{+0.005}_{-0.005}$ \\
  \multicolumn{2}{l}{$\mathcal{C}_\text{FPMA/DU1}$} & 1.371$^{+0.004}_{-0.004}$ & 1.098$^{+0.002}_{-0.002}$ & 1.136$^{+0.003}_{-0.004}$ \\
  \multicolumn{2}{l}{$\mathcal{C}_\text{FPMB/DU1}$} & 1.392$^{+0.005}_{-0.004}$ & 1.112$^{+0.002}_{-0.002}$ & 1.148$^{+0.004}_{-0.004}$ \\
  \multicolumn{2}{l}{$\mathcal{C}_\text{XTI/DU1}$} & / & 1.026$^{+0.002}_{-0.002}$ & 1.064$^{+0.005}_{-0.005}$ \\
\hline
\multicolumn{2}{l}{$\chi^2/\text{d.o.f.}$} & 549/526 & 350/309 & 318/275 \\
\hline
\multicolumn{5}{c}{Photon flux ratios (2--8 keV)} \\
\multicolumn{2}{l}{$N_\texttt{diskbb}/N_\text{Tot}$} & 24.6\% & 50.7\% & 24.1\% \\
\multicolumn{2}{l}{$N_\texttt{thcomp*bb}/N_\text{Tot}$} & 57.6\% & 33.5\% & 43.8\% \\
\multicolumn{2}{l}{$N_\texttt{relxillNS}/N_\text{Tot}$} & 17.8\% & 15.8\% & 32.1\% \\
\multicolumn{5}{c}{Photon flux ratios (2--4 keV)} \\
\multicolumn{2}{l}{$N_\texttt{diskbb}/N_\text{Tot}$} & 35.7\% & 61.5\% & 32.9\% \\
\multicolumn{2}{l}{$N_\texttt{thcomp*bb}/N_\text{Tot}$} & 46.3\% & 24.6\% & 35.4\% \\
\multicolumn{2}{l}{$N_\texttt{relxillNS}/N_\text{Tot}$} & 18.0\% & 13.9\% & 31.7\% \\
\multicolumn{5}{c}{Photon flux ratios (4--6 keV)} \\
\multicolumn{2}{l}{$N_\texttt{diskbb}/N_\text{Tot}$} & 6.5\% & 36.2\% & 10.3\% \\
\multicolumn{2}{l}{$N_\texttt{thcomp*bb}/N_\text{Tot}$} & 74.5\% & 45.4\% & 55.6\% \\
\multicolumn{2}{l}{$N_\texttt{relxillNS}/N_\text{Tot}$} & 19.0\% & 18.4\% & 34.1\% \\
\multicolumn{5}{c}{Photon flux ratios (6--8 keV)} \\
\multicolumn{2}{l}{$N_\texttt{diskbb}/N_\text{Tot}$} & 0.9\% & 17.7\% & 2.7\% \\
\multicolumn{2}{l}{$N_\texttt{thcomp*bb}/N_\text{Tot}$} & 85.3\% & 61.1\% & 67.6\% \\
\multicolumn{2}{l}{$N_\texttt{relxillNS}/N_\text{Tot}$} & 13.8\% & 21.2\% & 29.7\% \\
\multicolumn{5}{c}{Energy Flux (2--8 keV)} \\
  \multicolumn{2}{l}{$F_\text{Tot}$ ($10^{-9}$ \fluxcgs)} & 5.38 & 7.58 & 7.92 \\
\hline
\end{tabular}
\tablefoot{
Uncertainties are at the 90$\%$ confidence level for a single parameter. Parameters in common between the two columns were linked during the different branches, while those in square brackets were frozen during the fit. The normalizations of \diskbb and \bbodyrad are computed by assuming a source distance of 7.2 kpc \citep{Orosz.Kuulkers.1999}. $^{(\star)}$Best-fit taken by \cite{Gnarini.etAl.2026}.}
\end{table}

\begin{table}[h!]
\caption{Best-fitting model parameters for XTE~J1701--462.} 
\label{table:BestFit-XTEJ1701}      
\centering
\small
\begin{tabular}{@{}llccc@{}}
\noalign{\smallskip}
\hline\hline         
& Parameter & HB & NB & FB \\   
\hline     
\noalign{\smallskip}
\texttt{TBabs} & $N_{\rm H}$ ($10^{22}$\,cm$^{-2}$) & & 2.99$^{+0.08}_{-0.07}$ & \\
\texttt{edge} & $E_{\rm c}$ (keV) & [1.839] & / & / \\
 & $D$ ($10^{-2}$) & 5.8$^{+0.5}_{-0.5}$ & / & / \\
\texttt{diskbb} & $kT_\text{in}$ (keV) & 0.90$^{+0.02}_{-0.03}$ & 1.17$^{+0.06}_{-0.06}$ & 0.85$^{+0.02}_{-0.02}$ \\
 & $R_{\rm d} \sqrt{\cos i}$ (km) & 20.9$^{+0.6}_{-0.6}$ & 18.1$^{+1.8}_{-1.8}$ & 27.6$^{+1.1}_{-1.1}$ \\
 \texttt{thcomp} & $kT_{\rm e}$ (keV) & 2.8$^{+0.1}_{-0.1}$ & 2.8$^{+0.2}_{-0.2}$ & 3.0$^{+0.5}_{-0.5}$ \\
 & $\tau$ & 9.4$^{+0.8}_{-0.8}$ & 12.8$^{+2.5}_{-2.6}$ & 30.2$^{+9.3}_{-9.4}$ \\
 & $f$ & [1] & 0.19$^{+0.05}_{-0.05}$ & 0.021$^{+0.005}_{-0.006}$ \\
\texttt{bbodyrad} & $kT$ (keV) & 0.99$^{+0.03}_{-0.03}$ & 1.38$^{+0.01}_{-0.02}$ & 1.29$^{+0.01}_{-0.02}$ \\
& $R_\text{bb}$ (km) & 28.7$^{+2.0}_{-2.2}$ & 10.7$^{+1.7}_{-1.6}$ & 17.5$^{+0.3}_{-0.3}$ \\
\texttt{relxillNS} & $q_\text{em}$ & & [1.6] & \\
& $a$ & & [0.1] & \\
& $i$ (deg) & & 30$^{+6}_{-5}$ & \\
& $R_\text{in}$ ($R_{\rm ISCO}$) & \multicolumn{2}{c}{$<7.1$} & [1] \\
& $kT_\text{bb}$ (keV) & = $kT$ & = $kT$ & = $kT$ \\
& $\log \xi$ & \multicolumn{2}{c}{2.5$^{+0.3}_{-0.3}$} & 1.5$^{+0.4}_{-0.3}$ \\
& $A_\text{Fe}$ & & [1.5] & \\
& $\log n_{\rm e}$ (cm$^{-3}$) & & [18] & \\
& $N_{\rm r}$ ($10^{-3}$) & 1.4$^{+0.5}_{-0.4}$ & 3.3$^{+0.2}_{-0.1}$ & 3.3$^{+0.3}_{-0.4}$ \\
\hline
\multicolumn{5}{c}{Cross-calibration} \\
  \multicolumn{2}{l}{$\mathcal{C}_\text{DU2/DU1}$} & 1.002$^{+0.002}_{-0.002}$ & 1.003$^{+0.004}_{-0.004}$ & 1.010$^{+0.004}_{-0.003}$ \\
  \multicolumn{2}{l}{$\mathcal{C}_\text{DU3/DU1}$} & 0.969$^{+0.002}_{-0.002}$ & 0.976$^{+0.004}_{-0.004}$ & 0.985$^{+0.003}_{-0.003}$ \\
  \multicolumn{2}{l}{$\mathcal{C}_\text{FPMA/DU1}$} & / & 1.104$^{+0.003}_{-0.002}$ & 1.151$^{+0.002}_{-0.002}$ \\
  \multicolumn{2}{l}{$\mathcal{C}_\text{FPMB/DU1}$} & / & 1.115$^{+0.002}_{-0.002}$ & 1.152$^{+0.002}_{-0.002}$ \\
  \multicolumn{2}{l}{$\mathcal{C}_\text{XTI/DU1}$} & 1.158$^{+0.002}_{-0.002}$ & / & / \\
\hline
\multicolumn{2}{l}{$\chi^2/\text{d.o.f.}$} & 565/552 & 615/575 & 625/571 \\
\hline
\multicolumn{5}{c}{Photon flux ratios (2--8 keV)} \\
\multicolumn{2}{l}{$N_\texttt{diskbb}/N_\text{Tot}$} & 29.0\% & 67.3\% & 36.9\% \\
\multicolumn{2}{l}{$N_\texttt{thcomp*bb}/N_\text{Tot}$} & 67.5\% & 26.5\% & 61.4\% \\
\multicolumn{2}{l}{$N_\texttt{relxillNS}/N_\text{Tot}$} & 3.5\% & 6.2\% & 1.7\% \\
\multicolumn{5}{c}{Photon flux ratios (2--4 keV)} \\
\multicolumn{2}{l}{$N_\texttt{diskbb}/N_\text{Tot}$} & 38.4\% & 75.6\% & 48.1\% \\
\multicolumn{2}{l}{$N_\texttt{thcomp*bb}/N_\text{Tot}$} & 57.9\% & 19.6\% & 51.0\% \\
\multicolumn{2}{l}{$N_\texttt{relxillNS}/N_\text{Tot}$} & 3.7\% & 4.8\% & 0.9\% \\
\multicolumn{5}{c}{Photon flux ratios (4--6 keV)} \\
\multicolumn{2}{l}{$N_\texttt{diskbb}/N_\text{Tot}$} & 15.0\% & 53.9\% & 17.6\% \\
\multicolumn{2}{l}{$N_\texttt{thcomp*bb}/N_\text{Tot}$} & 81.6\% & 38.1\% & 79.8\% \\
\multicolumn{2}{l}{$N_\texttt{relxillNS}/N_\text{Tot}$} & 3.4\% & 8.0\% & 2.6\% \\
\multicolumn{5}{c}{Photon flux ratios (6--8 keV)} \\
\multicolumn{2}{l}{$N_\texttt{diskbb}/N_\text{Tot}$} & 4.9\% & 34.3\% & 5.7\% \\
\multicolumn{2}{l}{$N_\texttt{thcomp*bb}/N_\text{Tot}$} & 92.7\% & 52.6\% & 89.0\% \\
\multicolumn{2}{l}{$N_\texttt{relxillNS}/N_\text{Tot}$} & 2.4\% & 13.1\% & 5.7\% \\
\multicolumn{5}{c}{Energy Flux (2--8 keV)} \\
\multicolumn{2}{l}{$F_\text{Tot}$ ($10^{-8}$ \fluxcgs)} & 1.14 & 1.15 & 1.08 \\
\hline
\end{tabular}
\tablefoot{
Uncertainties are at the 90$\%$ confidence level for a single parameter. Parameters in common between the two columns were linked during the different branches, while those in square brackets were frozen during the fit. The normalizations of \diskbb and \bbodyrad are computed by assuming a source distance of 8.8 kpc \citep{Lin2007}.}
\end{table}

\begin{table}[h!]
\caption{Best-fitting model parameters for GX~5--1.} 
\label{table:BestFit-GX5-1}      
\centering                         
\small
\begin{tabular}{@{}llccc@{}}
\noalign{\smallskip}
\hline\hline         
& Parameter & HB & NB & FB \\   
\hline     
\noalign{\smallskip}
\texttt{TBabs} & $N_{\rm H}$ ($10^{22}$\,cm$^{-2}$) & & 4.54$^{+0.03}_{-0.03}$ & \\
\texttt{edge} & $E_{\rm c}$ (keV) & [1.839] & [1.839] & [1.839] \\
 & $D$ ($10^{-1}$) & 1.4$^{+0.3}_{-0.2}$ & 1.4$^{+0.3}_{-0.3}$ & 1.1$^{+0.1}_{-0.1}$ \\
\texttt{diskbb} & $kT_\text{in}$ (keV) & 1.18$^{+0.02}_{-0.03}$ & 1.05$^{+0.03}_{-0.03}$ & 1.19$^{+0.02}_{-0.02}$ \\
 & $R_{\rm d} \sqrt{\cos i}$ (km) & 16.4$^{+1.3}_{-1.3}$ & 20.6$^{+0.9}_{-0.8}$ & 18.4$^{+0.6}_{-0.6}$ \\
 \texttt{thcomp} & $kT_{\rm e}$ (keV) & 3.0$^{+0.1}_{-0.1}$ & 2.6$^{+0.1}_{-0.1}$ & / \\
 & $\tau$ & 8.3$^{+0.3}_{-0.4}$ & 7.5$^{+0.4}_{-0.4}$ & / \\
 & $f$ & [1] & [1] & [0] \\
\texttt{bbodyrad} & $kT$ (keV) & 1.47$^{+0.17}_{-0.14}$ & 1.11$^{+0.04}_{-0.03}$ & 1.57$^{+0.01}_{-0.01}$ \\
& $R_\text{bb}$ (km) & 13.6$^{+1.1}_{-1.0}$ & 24.4$^{+1.2}_{-1.3}$ & 12.2$^{+0.5}_{-0.5}$ \\
\texttt{expabs} & $E_\text{cut}$ & = $kT$ & = $kT$ & / \\
\texttt{powerlaw} & $\alpha$ & [2.62] & [2.62] & / \\
& $N_\text{pl}$ & 0.47$^{+0.04}_{-0.04}$ & 0.44$^{+0.05}_{-0.05}$ & [0] \\
\hline
\multicolumn{5}{c}{Cross-calibration} \\
  \multicolumn{2}{l}{$\mathcal{C}_\text{DU2/DU1}$} & 1.012$^{+0.002}_{-0.002}$ & 1.013$^{+0.003}_{-0.003}$ & 1.010$^{+0.004}_{-0.004}$ \\
  \multicolumn{2}{l}{$\mathcal{C}_\text{DU3/DU1}$} & 0.984$^{+0.002}_{-0.002}$ & 0.983$^{+0.002}_{-0.003}$ & 0.978$^{+0.004}_{-0.004}$ \\
  \multicolumn{2}{l}{$\mathcal{C}_\text{FPMA/DU1}$} & 1.324$^{+0.001}_{-0.001}$ & 1.228$^{+0.002}_{-0.002}$ & 1.234$^{+0.003}_{-0.002}$ \\
  \multicolumn{2}{l}{$\mathcal{C}_\text{FPMB/DU1}$} & 1.357$^{+0.001}_{-0.001}$ & 1.232$^{+0.002}_{-0.002}$ & 1.236$^{+0.002}_{-0.002}$ \\
  \multicolumn{2}{l}{$\mathcal{C}_\text{XTI/DU1}$} & 1.353$^{+0.001}_{-0.001}$ & 1.149$^{+0.002}_{-0.002}$ & 1.203$^{+0.002}_{-0.003}$ \\
\hline
\multicolumn{2}{l}{$\chi^2/\text{d.o.f.}$} & 976/881 & 906/814 & 871/708 \\
\hline
\multicolumn{5}{c}{Photon flux ratios (2--8 keV)} \\
\multicolumn{2}{l}{$N_\texttt{diskbb}/N_\text{Tot}$} & 52.8\% & 43.3\% & 60.0\% \\
\multicolumn{2}{l}{$N_\texttt{thcomp*bb}/N_\text{Tot}$} & 45.8\% & 55.3\% & 40.0\% \\
\multicolumn{2}{l}{$N_\texttt{powerlaw}/N_\text{Tot}$} & 1.4\% & 1.4\% & / \\
\multicolumn{5}{c}{Photon flux ratios (2--4 keV)} \\
\multicolumn{2}{l}{$N_\texttt{diskbb}/N_\text{Tot}$} & 65.5\% & 52.8\% & 70.6\% \\
\multicolumn{2}{l}{$N_\texttt{thcomp*bb}/N_\text{Tot}$} & 33.0\% & 45.8\% & 29.4\% \\
\multicolumn{2}{l}{$N_\texttt{powerlaw}/N_\text{Tot}$} & 1.5\% & 1.4\% & / \\
\multicolumn{5}{c}{Photon flux ratios (4--6 keV)} \\
\multicolumn{2}{l}{$N_\texttt{diskbb}/N_\text{Tot}$} & 38.1\% & 28.7\% & 45.3\% \\
\multicolumn{2}{l}{$N_\texttt{thcomp*bb}/N_\text{Tot}$} & 60.8\% & 70.2\% & 54.7\% \\
\multicolumn{2}{l}{$N_\texttt{powerlaw}/N_\text{Tot}$} & 1.1\% & 1.1\% & / \\
\multicolumn{5}{c}{Photon flux ratios (6--8 keV)} \\
\multicolumn{2}{l}{$N_\texttt{diskbb}/N_\text{Tot}$} & 18.8\% & 14.1\% & 26.7\% \\
\multicolumn{2}{l}{$N_\texttt{thcomp*bb}/N_\text{Tot}$} & 80.2\% & 84.7\% & 73.3\% \\
\multicolumn{2}{l}{$N_\texttt{powerlaw}/N_\text{Tot}$} & 1.0\% & 1.2\% & / \\
\multicolumn{5}{c}{Energy Flux (2--8 keV)} \\
  \multicolumn{2}{l}{$F_\text{Tot}$ ($10^{-8}$ \fluxcgs)} & 0.75 & 1.45 & 1.56 \\
\hline
\end{tabular}
\tablefoot{
Uncertainties are at the 90$\%$ confidence level for a single parameter. Parameters in common between the two columns were linked during the different branches, while those in square brackets were frozen during the fit. The normalizations of \diskbb and \bbodyrad are computed by assuming a source distance of 7.5 kpc \citep{Fabiani.etAl.2024}.}
\end{table}

\begin{table}[h!]
\caption{Best-fitting model parameters for Sco~X-1.} 
\label{table:BestFit-ScoX1}      
\centering                         
\small
\begin{tabular}{@{}llcc@{}}
\noalign{\smallskip}
\hline\hline         
& Parameter & NB & FB \\   
\hline     
\noalign{\smallskip}
\texttt{TBabs} & $N_{\rm H}$ ($10^{22}$\,cm$^{-2}$) & \multicolumn{2}{c}{$0.16^{+0.01}_{-0.01}$} \\
\texttt{edge} & $E_{\rm c}$ (keV) & \multicolumn{2}{c}{[1.839]} \\
 & $D$ ($10^{-2}$) & 3.6$^{+0.2}_{-0.2}$ & 2.7$^{+0.2}_{-0.2}$ \\
\texttt{diskbb} & $kT_\text{in}$ (keV) & 0.96$^{+0.01}_{-0.01}$ & 0.98$^{+0.01}_{-0.01}$ \\
 & $R_{\rm d} \sqrt{\cos i}$ (km) & 20.8$^{+0.6}_{-0.6}$ & 21.0$^{+0.4}_{-0.5}$ \\
\texttt{thcomp} & $kT_{\rm e}$ (keV) & 3.1$^{+0.1}_{-0.1}$ & 2.9$^{+0.2}_{-0.1}$ \\ 
 & $\tau$ & 8.0$^{+0.2}_{-0.3}$ & 11.1$^{+0.6}_{-0.5}$ \\
 & $f$ & [1] & 0.51$^{+0.03}_{-0.03}$ \\
\texttt{bbodyrad} & $kT$ (keV) & 1.31$^{+0.02}_{-0.02}$ & 1.50$^{+0.01}_{-0.02}$ \\
& $R_\text{bb}$ (km) & 14.1$^{+0.4}_{-0.4}$ & 14.3$^{+0.4}_{-0.3}$ \\
\texttt{relxillNS} & $q_\text{em}$ & \multicolumn{2}{c}{1.87$^{+0.05}_{-0.05}$} \\
& $a$ & \multicolumn{2}{c}{[0.1]} \\
& $i$ (deg) & \multicolumn{2}{c}{46$^{+3}_{-3}$} \\
& $R_\text{in}$ ($R_{\rm ISCO}$) & 2.2$^{+0.8}_{-0.8}$ & $<1.8$ \\
& $kT_\text{bb}$ (keV) & = $kT$ & = $kT$ \\
& $\log \xi$ & 2.9$^{+0.1}_{-0.1}$ & 2.5$^{+0.1}_{-0.2}$ \\
& $A_\text{Fe}$ & \multicolumn{2}{c}{3.0$^{+0.1}_{-0.1}$} \\
& $\log n_{\rm e}$ (cm$^{-3}$) & \multicolumn{2}{c}{[18]} \\
& $N_{\rm r}$ ($10^{-2}$) & 8.0$^{+0.3}_{-0.3}$ & 8.4$^{+0.4}_{-0.3}$ \\
\hline
\multicolumn{4}{c}{Cross-calibration} \\
  \multicolumn{2}{l}{$\mathcal{C}_\text{DU2/DU1}$} & 1.032$^{+0.002}_{-0.002}$ & 1.031$^{+0.002}_{-0.002}$ \\
  \multicolumn{2}{l}{$\mathcal{C}_\text{DU3/DU1}$} & 0.991$^{+0.002}_{-0.002}$ & 0.992$^{+0.002}_{-0.002}$ \\
  \multicolumn{2}{l}{$\mathcal{C}_\text{FPMA/DU1}$} & 0.801$^{+0.001}_{-0.001}$ & 0.836$^{+0.001}_{-0.001}$ \\
  \multicolumn{2}{l}{$\mathcal{C}_\text{FPMB/DU1}$} & 0.779$^{+0.001}_{-0.001}$ & 0.813$^{+0.001}_{-0.001}$ \\
  \multicolumn{2}{l}{$\mathcal{C}_\text{XTI/DU1}$} & 0.728$^{+0.002}_{-0.002}$ & 0.753$^{+0.002}_{-0.002}$ \\
\hline
\multicolumn{2}{l}{$\chi^2/\text{d.o.f.}$} & 440/367 & 474/367 \\
\hline
\multicolumn{4}{c}{Photon flux ratios (2--8 keV)} \\
\multicolumn{2}{l}{$N_\texttt{diskbb}/N_\text{Tot}$} & 42.3\% & 39.1\% \\
\multicolumn{2}{l}{$N_\texttt{thcomp*bb}/N_\text{Tot}$} & 44.6\% & 55.3\% \\
\multicolumn{2}{l}{$N_\texttt{relxillNS}/N_\text{Tot}$} & 13.1\% & 5.6\% \\
\multicolumn{4}{c}{Photon flux ratios (2--4 keV)} \\
\multicolumn{2}{l}{$N_\texttt{diskbb}/N_\text{Tot}$} & 55.1\% & 53.4\% \\
\multicolumn{2}{l}{$N_\texttt{thcomp*bb}/N_\text{Tot}$} & 33.5\% & 42.3\% \\
\multicolumn{2}{l}{$N_\texttt{relxillNS}/N_\text{Tot}$} & 11.4\% & 4.3\% \\
\multicolumn{4}{c}{Photon flux ratios (4--6 keV)} \\
\multicolumn{2}{l}{$N_\texttt{diskbb}/N_\text{Tot}$} & 24.4\% & 21.9\% \\
\multicolumn{2}{l}{$N_\texttt{thcomp*bb}/N_\text{Tot}$} & 60.0\% & 71.4\% \\
\multicolumn{2}{l}{$N_\texttt{relxillNS}/N_\text{Tot}$} & 15.6\% & 6.7\% \\
\multicolumn{4}{c}{Photon flux ratios (6--8 keV)} \\
\multicolumn{2}{l}{$N_\texttt{diskbb}/N_\text{Tot}$} & 8.4\% & 7.3\% \\
\multicolumn{2}{l}{$N_\texttt{thcomp*bb}/N_\text{Tot}$} & 74.5\% & 82.9\% \\
\multicolumn{2}{l}{$N_\texttt{relxillNS}/N_\text{Tot}$} & 17.1\% & 9.8\% \\
\multicolumn{4}{c}{Energy Flux (2--8 keV)} \\
\multicolumn{2}{l}{$F_\text{Tot}$ ($10^{-7}$ \fluxcgs)} & 2.16 & 2.83 \\
\hline
\end{tabular}
\tablefoot{
Uncertainties are at the 90$\%$ confidence level for a single parameter. Parameters in common between the two columns were linked during the different branches, while those in square brackets were frozen during the fit. The normalizations of \diskbb and \bbodyrad are computed by assuming a source distance of 2.1 kpc \citep{Arnason.etAl.2021}.}
\end{table}

\begin{table}[h!]
\caption{Best-fitting model parameters for GX~340+0.} 
\label{table:BestFit-GX340+0}      
\centering                         
\small
\begin{tabular}{@{}llccc@{}}
\noalign{\smallskip}
\hline\hline         
& Parameter & HB & NB & FB \\   
\hline     
\noalign{\smallskip}
\texttt{TBabs} & $N_{\rm H}$ ($10^{22}$\,cm$^{-2}$) & & 8.86$^{+0.06}_{-0.07}$ & \\
\texttt{edge} & $E_{\rm c}$ (keV) & / & [1.839] & [1.839] \\
 & $D$ ($10^{-1}$) & / & 2.3$^{+0.5}_{-0.5}$ & 1.5$^{+0.3}_{-0.3}$ \\
\texttt{diskbb} & $kT_\text{in}$ (keV) & 1.15$^{+0.05}_{-0.05}$ & 1.26$^{+0.07}_{-0.06}$ & 1.04$^{+0.08}_{-0.08}$ \\
 & $R_{\rm d} \sqrt{\cos i}$ (km) & 18.5$^{+1.5}_{-1.5}$ & 17.6$^{+1.2}_{-1.1}$ & 23.2$^{+1.3}_{-1.2}$ \\
 \texttt{thcomp} & $kT_{\rm e}$ (keV) & 3.3$^{+0.3}_{-0.3}$ & 2.9$^{+0.3}_{-0.3}$ & 14.5$^{+1.5}_{-1.5}$ \\
 & $\tau$ & 7.7$^{+0.5}_{-0.6}$ & 10.1$^{+0.7}_{-0.6}$ & 21.3$^{+4.3}_{-4.5}$ \\
 & $f$ & [1] & 0.34$^{+0.05}_{-0.04}$ & 0.010$^{+0.003}_{-0.003}$ \\
\texttt{bbodyrad} & $kT$ (keV) & 1.49$^{+0.08}_{-0.08}$ & 1.47$^{+0.05}_{-0.05}$ & 1.43$^{+0.05}_{-0.05}$ \\
& $R_\text{bb}$ (km) & 14.0$^{+1.7}_{-1.7}$ & 12.5$^{+2.0}_{-2.1}$ & 15.6$^{+0.7}_{-0.7}$ \\
\texttt{relxillNS} & $q_\text{em}$ & \multicolumn{2}{c}{[2.2]} & [2.4] \\
& $a$ & & [0.1] & \\
& $i$ (deg) & & 36$^{+4}_{-5}$ & \\
& $R_\text{in}$ ($R_{\rm ISCO}$) & $<3.4$ & $<2.0$ & $<3.2$ \\
& $kT_\text{bb}$ (keV) & = $kT$ & = $kT$ & = $kT$ \\
& $\log \xi$ & 2.7$^{+0.1}_{-0.1}$ & 2.5$^{+0.3}_{-0.3}$ & 2.4$^{+0.1}_{-0.1}$ \\
& $A_\text{Fe}$ & & [9.7] & \\
& $\log n_{\rm e}$ (cm$^{-3}$) & & [18] & \\
& $N_{\rm r}$ ($10^{-3}$) & 1.4$^{+0.2}_{-0.2}$ & 1.6$^{+0.6}_{-0.5}$ & 2.3$^{+0.2}_{-0.3}$ \\
\texttt{apec} & $kT_\text{a}$ (keV) & \multicolumn{2}{c}{[2.24]} & [2.10] \\
& $A_{\rm Z}$ & & [2.26] & \\
& $N_\text{apec}$ ($10^{-2}$) & $<2.4$ & 4.2$^{+0.3}_{-0.3}$ & 7.9$^{+0.6}_{-0.6}$ \\
\hline
\multicolumn{5}{c}{Cross-calibration} \\
  \multicolumn{2}{l}{$\mathcal{C}_\text{DU2/DU1}$} & 1.007$^{+0.003}_{-0.003}$ & 1.005$^{+0.002}_{-0.002}$ & 1.012$^{+0.006}_{-0.006}$ \\
  \multicolumn{2}{l}{$\mathcal{C}_\text{DU3/DU1}$} & 0.982$^{+0.002}_{-0.002}$ & 0.983$^{+0.002}_{-0.002}$ & 0.989$^{+0.006}_{-0.006}$ \\
  \multicolumn{2}{l}{$\mathcal{C}_\text{FPMA/DU1}$} & 1.275$^{+0.001}_{-0.001}$ & 1.142$^{+0.002}_{-0.002}$ & 1.149$^{+0.003}_{-0.003}$ \\
  \multicolumn{2}{l}{$\mathcal{C}_\text{FPMB/DU1}$} & 1.248$^{+0.001}_{-0.001}$ & 1.121$^{+0.002}_{-0.002}$ & 1.125$^{+0.004}_{-0.004}$ \\
  \multicolumn{2}{l}{$\mathcal{C}_\text{XTI/DU1}$} & / & 1.097$^{+0.002}_{-0.002}$ & 1.011$^{+0.005}_{-0.005}$ \\
\hline
\multicolumn{2}{l}{$\chi^2/\text{d.o.f.}$} & 763/675 & 808/750 & 785/682 \\
\hline
\multicolumn{5}{c}{Photon flux ratios (2--8 keV)} \\
\multicolumn{2}{l}{$N_\texttt{diskbb}/N_\text{Tot}$} & 53.3\% & 66.7\% & 50.8\% \\
\multicolumn{2}{l}{$N_\texttt{thcomp*bb}/N_\text{Tot}$} & 44.4\% & 32.0\% & 47.1\% \\
\multicolumn{2}{l}{$N_\texttt{relxillNS}/N_\text{Tot}$} & 2.3\% & 1.3\% & 2.1\% \\
\multicolumn{5}{c}{Photon flux ratios (2--4 keV)} \\
\multicolumn{2}{l}{$N_\texttt{diskbb}/N_\text{Tot}$} & 66.6\% & 75.7\% & 61.8\% \\
\multicolumn{2}{l}{$N_\texttt{thcomp*bb}/N_\text{Tot}$} & 31.7\% & 23.3\% & 36.5\% \\
\multicolumn{2}{l}{$N_\texttt{relxillNS}/N_\text{Tot}$} & 1.7\% & 1.0\% & 1.7\% \\
\multicolumn{5}{c}{Photon flux ratios (4--6 keV)} \\
\multicolumn{2}{l}{$N_\texttt{diskbb}/N_\text{Tot}$} & 37.6\% & 54.3\% & 33.8\% \\
\multicolumn{2}{l}{$N_\texttt{thcomp*bb}/N_\text{Tot}$} & 59.6\% & 44.2\% & 64.2\% \\
\multicolumn{2}{l}{$N_\texttt{relxillNS}/N_\text{Tot}$} & 2.8\% & 1.5\% & 2.0\% \\
\multicolumn{5}{c}{Photon flux ratios (6--8 keV)} \\
\multicolumn{2}{l}{$N_\texttt{diskbb}/N_\text{Tot}$} & 17.3\% & 35.5\% & 16.5\% \\
\multicolumn{2}{l}{$N_\texttt{thcomp*bb}/N_\text{Tot}$} & 78.0\% & 61.0\% & 79.0\% \\
\multicolumn{2}{l}{$N_\texttt{relxillNS}/N_\text{Tot}$} & 4.7\% & 3.5\% & 4.5\% \\
\multicolumn{5}{c}{Energy Flux (2--8 keV)} \\
\multicolumn{2}{l}{$F_\text{Tot}$ ($10^{-9}$ \fluxcgs)} & 5.72 & 6.58 & 6.29 \\
\hline
\end{tabular}
\tablefoot{
Uncertainties are at the 90$\%$ confidence level for a single parameter. Parameters in common between the two columns were linked during the different branches, while those in square brackets were frozen during the fit. The normalizations of \diskbb and \bbodyrad are computed by assuming a source distance of 11 kpc \citep{Fender.Hendry.2000}.}
\end{table}

\begin{table}[h!]
\caption{Best-fitting model parameters for GX~349+2.} 
\label{table:BestFit-GX349+2}      
\centering                         
\small
\begin{tabular}{ll cc}        
\noalign{\smallskip}
\hline\hline         
& Parameter & NB & FB \\   
\hline     
\noalign{\smallskip}
\texttt{TBabs} & $N_{\rm H}$ ($10^{22}$\,cm$^{-2}$) & \multicolumn{2}{c}{1.38$^{+0.04}_{-0.04}$} \\
\texttt{diskbb} & $kT_\text{in}$ (keV) & 0.96$^{+0.02}_{-0.02}$ & 0.80$^{+0.03}_{-0.03}$ \\
 & $R_{\rm d} \sqrt{\cos i}$ (km) & 21.7$^{+0.7}_{-0.8}$ & 29.5$^{+1.6}_{-1.5}$ \\
 \texttt{thcomp} & $kT_{\rm e}$ (keV) & 3.0$^{+0.1}_{-0.1}$ & 2.9$^{+0.1}_{-0.1}$ \\
 & $\tau$ & 9.2$^{+0.7}_{-0.6}$ & 8.4$^{+1.0}_{-0.9}$ \\
 & $f$ & [1] & [1] \\
\texttt{bbodyrad} & $kT$ (keV) & 1.39$^{+0.03}_{-0.02}$ & 1.27$^{+0.03}_{-0.03}$ \\
& $R_\text{bb}$ (km) & 13.9$^{+0.5}_{-0.5}$ & 16.5$^{+0.9}_{-0.8}$ \\
\texttt{relxillNS} & $q_\text{em}$ & \multicolumn{2}{c}{[3]} \\
& $a$ & \multicolumn{2}{c}{[0.1]} \\
& $i$ (deg) & \multicolumn{2}{c}{37$^{+5}_{-5}$} \\
& $R_\text{in}$ ($R_{\rm ISCO}$) & $ < 1.6$& [1] \\
& $kT_\text{bb}$ (keV) & = $kT$ & = $kT$ \\
& $\log \xi$ & 3.0$^{+0.1}_{-0.1}$ & 2.9$^{+0.1}_{-0.2}$ \\
& $A_\text{Fe}$ & \multicolumn{2}{c}{5.8$^{+0.4}_{-0.4}$} \\
& $\log n_{\rm e}$ (cm$^{-3}$) & \multicolumn{2}{c}{[18]} \\
& $N_{\rm r}$ ($10^{-3}$) & 3.4$^{+0.2}_{-0.2}$ & 7.3$^{+0.1}_{-0.1}$ \\
\hline
\multicolumn{4}{c}{Cross-calibration} \\
  \multicolumn{2}{l}{$\mathcal{C}_\text{DU2/DU1}$} & 1.001$^{+0.002}_{-0.002}$ & 1.003$^{+0.002}_{-0.002}$ \\
  \multicolumn{2}{l}{$\mathcal{C}_\text{DU3/DU1}$} & 0.976$^{+0.002}_{-0.002}$ & 0.979$^{+0.002}_{-0.002}$ \\
  \multicolumn{2}{l}{$\mathcal{C}_\text{FPMA/DU1}$} & 1.149$^{+0.001}_{-0.001}$ & 1.147$^{+0.002}_{-0.001}$ \\
  \multicolumn{2}{l}{$\mathcal{C}_\text{FPMB/DU1}$} & 1.121$^{+0.001}_{-0.001}$ & 1.122$^{+0.001}_{-0.002}$ \\
\hline
\multicolumn{2}{l}{$\chi^2/\text{d.o.f.}$} & 683/647 & 657/632 \\
\hline
\multicolumn{4}{c}{Photon flux ratios (2--8 keV)} \\
\multicolumn{2}{l}{$N_\texttt{diskbb}/N_\text{Tot}$} & 43.6\% & 32.5\% \\
\multicolumn{2}{l}{$N_\texttt{thcomp*bb}/N_\text{Tot}$} & 47.3\% & 50.8\% \\
\multicolumn{2}{l}{$N_\texttt{relxillNS}/N_\text{Tot}$} & 9.1\% & 16.7\% \\
\multicolumn{4}{c}{Photon flux ratios (2--4 keV)} \\
\multicolumn{2}{l}{$N_\texttt{diskbb}/N_\text{Tot}$} & 57.7\% & 45.5\% \\
\multicolumn{2}{l}{$N_\texttt{thcomp*bb}/N_\text{Tot}$} & 34.6\% & 39.3\% \\
\multicolumn{2}{l}{$N_\texttt{relxillNS}/N_\text{Tot}$} & 7.7\% & 15.2\% \\
\multicolumn{4}{c}{Photon flux ratios (4--6 keV)} \\
\multicolumn{2}{l}{$N_\texttt{diskbb}/N_\text{Tot}$} & 24.9\% & 13.6\% \\
\multicolumn{2}{l}{$N_\texttt{thcomp*bb}/N_\text{Tot}$} & 63.8\% & 66.8\% \\
\multicolumn{2}{l}{$N_\texttt{relxillNS}/N_\text{Tot}$} & 11.3\% & 19.6\% \\
\multicolumn{4}{c}{Photon flux ratios (6--8 keV)} \\
\multicolumn{2}{l}{$N_\texttt{diskbb}/N_\text{Tot}$} & 8.3\% & 3.1\% \\
\multicolumn{2}{l}{$N_\texttt{thcomp*bb}/N_\text{Tot}$} & 80.0\% & 78.3\% \\
\multicolumn{2}{l}{$N_\texttt{relxillNS}/N_\text{Tot}$} & 11.7\% & 18.6\% \\
\multicolumn{4}{c}{Energy Flux (2--8 keV)} \\
\multicolumn{2}{l}{$F_\text{Tot}$ ($10^{-8}$ \fluxcgs)} & 1.03 & 1.04 \\
\hline
\end{tabular}
\tablefoot{
Uncertainties are at the 90$\%$ confidence level for a single parameter. Parameters in common between the two columns were linked during the different branches, while those in square brackets were frozen during the fit. The normalizations of \diskbb and \bbodyrad are computed by assuming a source distance of 9 kpc \citep{Grimm.etAl.2002}.}
\end{table}

\end{appendix}

\end{document}